\documentclass[aps, preprint, superscriptaddress]{revtex4-2}
\usepackage{graphicx}
\usepackage{makecell, multirow}
\usepackage{color, bm, soul, xcolor}
\usepackage{amsmath, amssymb, amsfonts}
\usepackage{gensymb}
\usepackage{array}
\usepackage{dcolumn}
\usepackage{geometry}
\usepackage{braket}
\usepackage{diagbox}
\usepackage[colorlinks=true, linkcolor=blue, citecolor=blue, urlcolor=blue]{hyperref}
\begin{document}
\renewcommand{\abstractname}{}

\title{Rippled Moir\'e Superlattices for Decoupled Ferroelectric Bits}

\author{Di Fan}
\affiliation{Department of Physics, School of Science, Westlake University, Hangzhou, Zhejiang 310030, China}

\author{Changming Ke}
\email{kechangming@westlake.edu.cn}
\affiliation{Department of Physics, School of Science, Westlake University, Hangzhou, Zhejiang 310030, China}
\affiliation{Institute of Natural Sciences, Westlake Institute for Advanced Study, Hangzhou, Zhejiang 310024, China}

\author{Shi Liu}
\email{liushi@westlake.edu.cn}
\affiliation{Department of Physics, School of Science, Westlake University, Hangzhou, Zhejiang 310030, China}
\affiliation{Institute of Natural Sciences, Westlake Institute for Advanced Study, Hangzhou, Zhejiang 310024, China}

\begin{abstract}
Symmetry considerations suggest that moir\'e superlattices formed by twisted two-dimensional materials should  preserve overall inversion symmetry. However, experiments consistently report robust ferroelectricity in systems such as twisted bilayer $h$-BN, posing a fundamental discrepancy between theory and experiment regarding its microscopic origin.
Here, using large-scale finite-field molecular dynamics simulations, we challenge the prevailing defect-pinning hypothesis and instead identify an out-of-plane bending field, induced by in-plane compressive strain, as the key symmetry-breaking mechanism. This strain-induced rippling drives spatially heterogeneous interlayer sliding and distorts the moir\'e domain wall network, resulting in a four-state ferroelectric system.
Remarkably, we show this mechanism can be harnessed at the nanoscale, where localized nanobubbles designate the moir\'e lattice's fundamental hexagonal domain clusters as the smallest individually addressable ferroelectric bits, thereby imposing local control on an otherwise globally defined structure.
Our findings establish a geometry-driven framework for understanding and engineering moir\'e ferroelectrics, offering not only a route toward ultra-high-density, rewritable memory, but also a strategy for locally tuning the moir\'e potential itself, a critical step for manipulating emergent correlated and topological quantum phases.

\end{abstract}

\maketitle

\clearpage
The discovery of sliding ferroelectricity, where switchable polarization arises from the relative displacement of layers in van der Waals (vdW) materials,  has opened new frontiers in the study of two-dimensional (2D) materials~\cite{ViznerStern21p1462,Yasuda21p6549,Wu21pe2115703118, Li17p6382,Gao24p196801}. A particularly rich manifestation of this phenomenon appears in moir\'e ferroelectricity, which emerges when vdW layers are rotated by a small twist angle. This twist creates a moir\'e superlattice, inducing a periodic modulation of local atomic stacking that generates an ordered array of interfacial dipoles.
Moir\'e ferroelectricity has been experimentally observed across a diverse range of systems, including twisted bilayer and trilayer transition metal dichalcogenides (TMDs)~\cite{Wang22p367, Weston22p390,VanWinkle24p751}, twisted bilayer hexagonal boron nitride ($h$-BN)~\cite{ViznerStern21p1462, Yasuda21p6549,guo25p11543}, twisted double bilayer graphene~\cite{du24p025015} and various vdW heterostructures~\cite{Zheng20p71,niu22p6241}. Notably, moir\'e superlattices are also known to host a rich landscape of correlated and topological quantum phases~\cite{Cao18p43, can21p519, song19p123, Lian20p12, serlin20p367}. The coexistence of ferroelectricity in these systems introduces a new degree of control, where ferroelectric polarization could potentially serve as an intrinsic, nonvolatile gate, enabling electrical modulation of quantum states and providing a versatile platform for reconfigurable quantum devices~\cite{Wu23p4,kang23p861}.

\begin{figure}[t]
\includegraphics[width=0.85\textwidth]{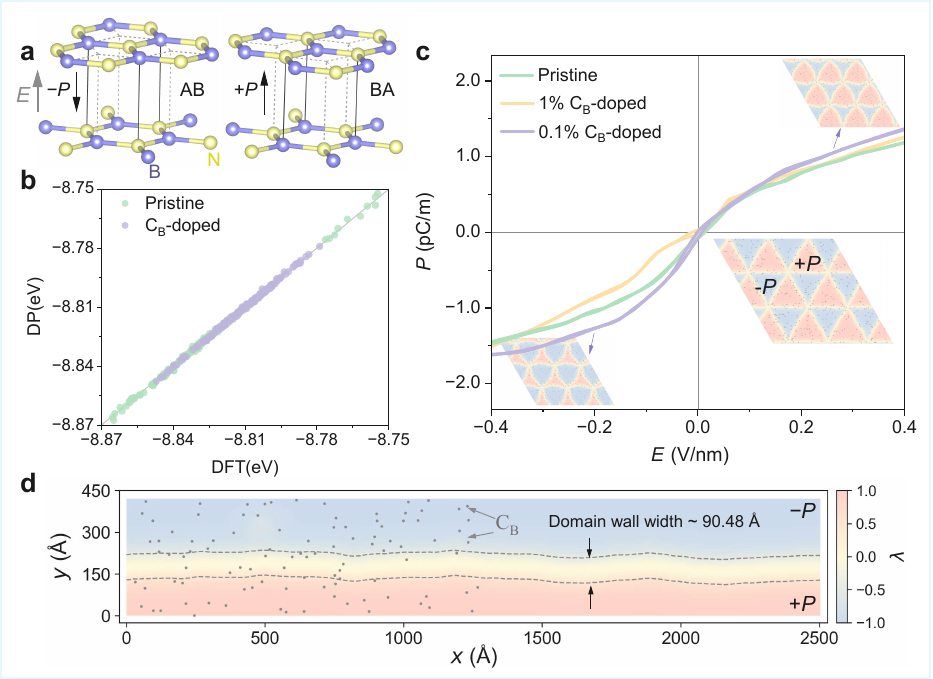}
\caption{\textbf{Effect of point defects on polarization and domain walls in twisted bilayer $h$-BN.} (\textbf{a}) Schematics of AB and BA stacking domains in twisted bilayer $h$-BN. (\textbf{b}) Comparison of energies computed with DFT and DP model for pristine and C$_\mathrm{B}$-doped configurations in the training database.  (\textbf{c}) Polarization--electric field ($P$--$E$) hysteresis loops for pristine twisted bilayer (with twist angle $\theta=0.365$\degree) and systems with 0.1\% and 1\% C$_\mathrm{B}$ defects. All cases exhibit a non-hysteretic response with zero remanent polarization, indicating the absence of defect-induced ferroelectricity. Insets show domain patterns for a moir\'e superlattice with 0.1\% C$_\mathrm{B}$ defects (black gray dots) at zero field. 
(\textbf{d}) Equilibrium structure of an extended $\Sigma_0$ 180~\degree~domain wall~\cite{ke25p046201} in Bernal-stacked bilayer $h$-BN modeled using a 771840-atom supercell at 300 K. The supercell spans approximately $2500~\text{\AA} \times 420~\text{\AA}$ in the in-plane directions, with a defect-rich region on the left and a pristine region on the right. The domain wall propagates continuously across both regions without noticeable distortion or bending. Color indicates the local structural order parameter ($\lambda$)~\cite{ke25p046201}. }
\label{fig:defect}
\end{figure}

The emergence of moir\'e ferroelectricity is surprising from a symmetry standpoint. In moir\'e superlattices, the periodic variation in atomic stacking leads to domains, such as AB and BA stacking (Fig.~\ref{fig:defect}a) in bilayer $h$-BN, that are structurally distinct yet energetically equivalent. By symmetry, these domains should form in equal areas~\cite{Cao18p80, guo25p11543, Weston20p592}. If they carry opposing physical properties, their effects should cancel, resulting in an overall non-polar or antiferroelectric-like ground state. Take twisted bilayer $h$-BN as an example, the moir\'e superlattice exhibits triangular tessellation of AB and BA domains with opposite out-of-plane polarizations~\cite{Bennett22p235445,moore21p5741,Dang25p085102,He24p119416}, which are theoretically expected to yield zero net polarization. This picture is supported by experiments in which polarization is weak or disappears entirely when the external electric field is removed~\cite{VanWinkle24p751, Ko23p992}.

However, this symmetry-based expectation is strongly challenged by experiments reporting robust ferroelectricity, including clear hysteresis loops and substantial remanent polarization~\cite{ViznerStern21p1462, Wang22p367, Weston22p390, VanWinkle24p751, Yasuda21p6549,guo25p11543, Yasuda21p6549, wong25p2414442, fan25p3557, ming22p21}. Notably, the observed ferroelectric states are often linked to asymmetric AB and BA domains, manifesting as contracted or expanded triangular regions bounded by bent domain walls. This discrepancy points to a missing ingredient: a symmetry-breaking mechanism that stabilizes a net polarization. Uncovering this mechanism is crucial, as it governs not only the emergence of ferroelectricity but is also highly relevant to the precise control of the moir\'e potential, which in turn sensitively modulates the correlated electronic states hosted within the superlattice.

To elucidate the origin of moir\'e ferroelectricity, we investigate two extrinsic factors that are unavoidable in realistic twisted bilayer $h$-BN samples: point defects and mechanical strain. Point defects are a natural candidate, as their role in pinning domains and domain walls is well-established in conventional ferroelectrics~\cite{Tybell02p097601,Li04p5261,philippe21p117601}. It has thus been proposed that they could be responsible for stabilizing a net polarization in moir\'e systems~\cite{li25p5451,He24p119416}. Here, we focus on carbon substitutions at boron sites (C$_\mathrm{B}$), a prevalent impurity in fabricated $h$-BN~\cite{weston20p214104,uddin17p18}, as a representative defect. Mechanical strain is also inherent to sample fabrication, arising from lattice or thermal expansion mismatches between layers and interactions with the substrate~\cite{kim23p2107362}. The strain-induced stresses often relax into out-of-plane deformations such as ripples, wrinkles, and localized bubbles~\cite{Blundo22p1525,wang19p116101}, which can fundamentally alter the interlayer geometry and break structural symmetries.  To investigate these phenomena at realistic scales, we developed a first-principles-accurate machine-learning interatomic potential for the C$_\mathrm{B}$-doped $h$-BN system within the deep potential (DP) framework~\cite{Zhang18p143001,Zhang18p4441}. 
As shown in Fig.~\ref{fig:defect}b, the DP model demonstrates excellent agreement with DFT energies for both pristine and C$_\mathrm{B}$-doped configurations, achieving a mean absolute error (MAE) of 0.413 meV/atom (See the Supplementary Materials for additional validation).
This model enables large-scale, finite-temperature molecular dynamics (MD) simulations that resolve the response of twisted $h$-BN to point defects and strain at the atomic scale.

Our simulations reveal that point defects, contrary to common expectation, are surprisingly ineffective at inducing stable ferroelectricity. Instead, we identify compressive strain as the fundamental driving force, which induces out-of-plane buckling that breaks the symmetry between AB and BA domains, leading to a net polarization. A striking consequence of this mechanism is the emergence of a four-state ferroelectric system governed by two coupled order parameters, the buckling amplitude and the AB/BA domain ratio. We further demonstrate that this strain-driven mechanism can be harnessed at the nanoscale using localized nanobubbles, which act as natural strain concentrators such that moir\'e superlattice containing arrays of nanobubbles display robust, symmetric ferroelectric switching. Remarkably, each nanobubble renders the surrounding hexagonal domain cluster an individually addressable ferroelectric bit. This discovery opens a pathway toward ultra-dense memory device architectures composed of programmable ferroelectric bits embedded in moir\'e superlattices.

We first examine the hypothesis that point defects are responsible for stabilizing moir\'e ferroelectricity. To establish a baseline, we confirm that an ideal, defect-free bilayer $h$-BN with a twist angle $\theta = 0.365^\circ$ (modeled with a 98284-atom supercell) relaxes into a moir\'e superlattice of alternating AB and BA triangular domains with nearly equal areas, forming a globally nonpolar ground state (Fig.~S3).
We then introduce C$_\mathrm{B}$ defects into this system and perform large-scale finite-field MD simulations to map the polarization–electric field ($P$--$E$) hysteresis loops, with $E$ applied along the out-of-plane direction. These simulations employ a dynamic, environment-dependent Born effective charge model from our previous work~\cite{ke25p046201}. This model is critical for calculating the evolving electrostatic driving forces induced by the applied field on the fly, as it continuously accounts for the charge transfer associated with structural changes during interlayer sliding.
As shown in Fig.~\ref{fig:defect}c, even with a high concentration of C$_\mathrm{B}$ defects (up to 1\%), the system exhibits a non-hysteretic response, nearly identical to the defect-free case. Upon removal of the external field, the AB and BA domains recover their nearly equal areas, resulting in zero remanent polarization. This demonstrates that point defects like C$_\mathrm{B}$ are insufficient to break the global antiferroelectric symmetry.

For a net polarization to emerge, the balance between AB and BA domains must be broken. Structurally, this typically requires bent domain walls (DWs) that expand one domain type at the expense of the other. The observed lack of remanent polarization in our simulations thus implies that point defects in twisted $h$-BN fail to pin DWs, a stark contrast to conventional perovskites where defect pinning is a defining feature~\cite{dilshood24p132901,dr20p033159,bencan20p1762}. 
A direct comparison of DW morphology with and without defects confirms the absence of pinning.
Specifically, simulations of a 2500 \AA-long DW show it traversing both pristine and defect-rich regions without any noticeable distortion, bending, or other evidence of pinning (Fig.~\ref{fig:defect}d). 
We attribute the absence of defect pinning to the intrinsically broad nature of moir\'e DWs.  Unlike the atomically sharp DWs in perovskites, those in twisted bilayer $h$-BN are delocalized over tens of nanometers~\cite{dong25pL201406,ke25p046201,xu25parXiv,shi25p035421}. This spatial broadening effectively dilutes the influence of individual point defects, rendering their energetic contribution negligible relative to the total domain wall energy. Consequently, defects are unable to exert sufficient pinning forces to immobilize the wall or disrupt the AB/BA domain symmetry.
This principle not only explains why point defects cannot induce ferroelectricity but also highlights a key characteristic of  moir\'e systems: their switching mechanism is fundamentally resilient to local disorder, consistent with the experimentally observed fatigue-free behavior~\cite{Bian24p57,Yasuda24peadp3575}.

\begin{figure}[t]
\includegraphics[width=0.95\textwidth]{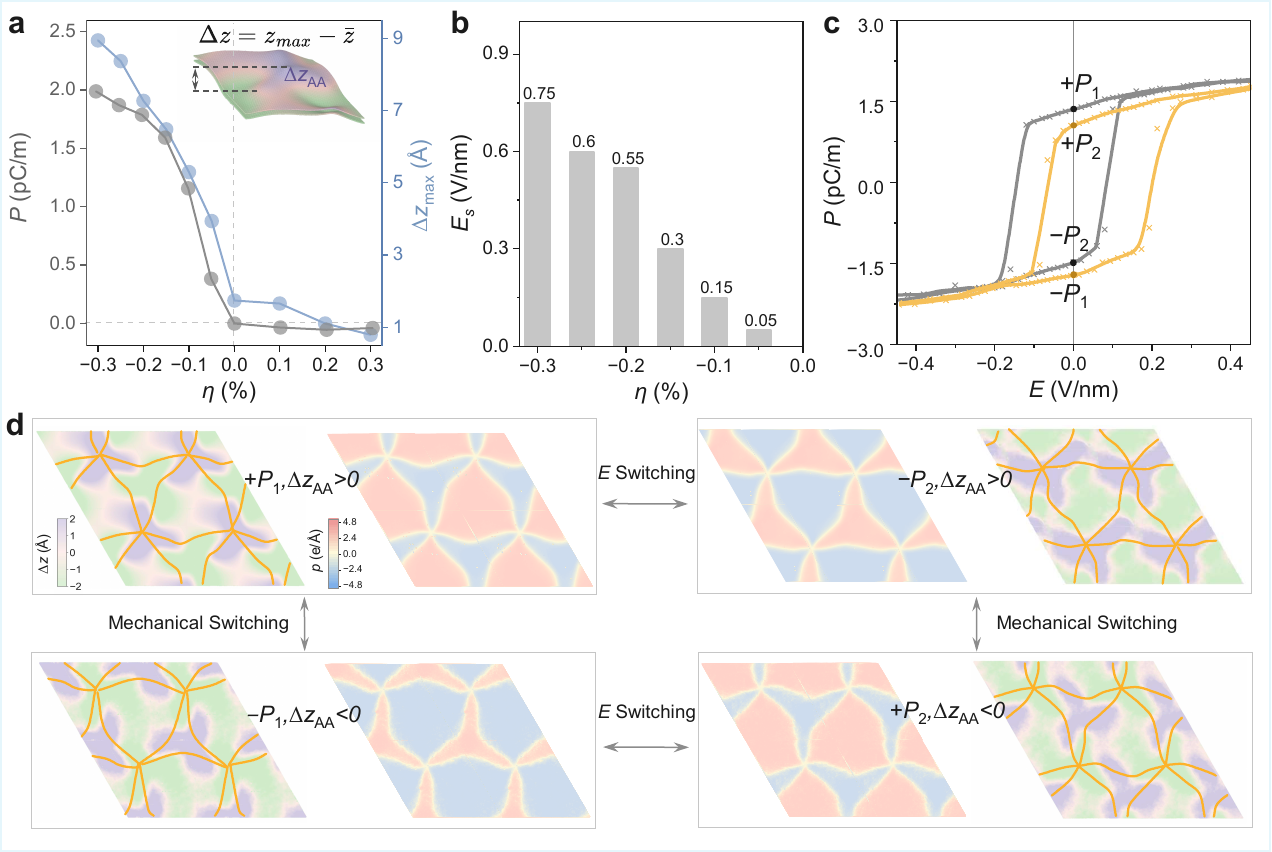}
\caption{\textbf{Strain-induced moir\'e ferroelectricity in twisted bilayer $h$-BN.}  
(\textbf{a}) Net out-of-plane polarization ($P$) and buckling amplitude ($\Delta z_{\rm max}$) as functions of in-plane biaxial strain ($\eta$).
(\textbf{b}) Switching field ($E_s$) as a function of $\eta$.
(\textbf{c}) $P$--$E$ hysteresis loops at $-0.1\%$ strain for moir\'e superlattices with upward ($\Delta z_{\rm AA} > 0$, black) and downward ($\Delta z_{\rm AA} < 0$, yellow) AA buckling.
(\textbf{d}) Local polarization ($p$) and buckling ($\Delta z$) profiles for the four polar states. 
The orange lines in the $\Delta z$-profiles indicate the locations of domain walls.
Electrical switching enables reversible transitions of $+P_1\leftrightarrow-P_2$ and $+P_2\leftrightarrow-P_1$ within a fixed buckling state. Mechanical switching reverses the sign of $\Delta z_{\rm AA}$, enabling transitions between $+P_1\leftrightarrow -P_1$ and $+P_2\leftrightarrow-P_2$.}
\label{fig:strain}
\end{figure}

We next examine the effect of in-plane strain ($\eta$) on the structure of the moir\'e superlattice. To quantify the out-of-plane structural corrugation, we introduce the buckling amplitude, $\Delta z_{\rm max}$, defined as the maximum vertical displacement ($z_{\rm max}$) relative to the geometric center ($\bar{z}$) of the bilayer. We note that even in the fully relaxed state ($\eta = 0$), the system exhibits mild buckling ($z_{\rm max}=$3.48~\AA), with the AA stacking regions typically protruding outward the most (Fig.~\ref{fig:strain}a inset). This behavior arises because the AA stacking configuration is significantly less stable than the AB and BA regions (see Fig.~S4), making out-of-plane buckling in the AA regions energetically more favorable.
The direction of this buckling, either upward ($\Delta z_{\rm AA} > 0$) or downward ($\Delta z_{\rm AA} < 0$), is actually an additional, and often underappreciated, order parameter characterizing the polar state of moir\'e superlattice (see discussion below).

Applying tensile strain ($\eta > 0$) flattens the bilayer and reduces $\Delta z_{\rm max}$. Importantly, we find that even a small compressive strain of $-0.05\%$ significantly increases $\Delta z_{\rm max}$ and induces a net polarization of 0.38 pC/m. 
Further increasing the compressive strain enhances both the buckling amplitude and the polarization magnitude (Fig.~\ref{fig:strain}a). 
This net polarization arises from the bending of DWs, which expand the AB domain area relative to the BA area, as visualized in Fig.~\ref{fig:strain}d (top left).
To evaluate the switchability of this compressive strain-induced polarization, we define the switching field ($E_s$) as the minimum electric field required to reverse the polarization within 200 ps in MD simulations. As shown in Fig.~\ref{fig:strain}b, $E_s$ increases with compressive strain, indicating that stronger remanent polarization is accompanied by higher coercivity. These results reveal that compressive strain plays a crucial role in activating and stabilizing ferroelectricity in the moir\'e superlattice.

Interestingly, the interplay between AB/BA domain ratio and AA buckling ($\Delta z_{\rm AA}$) gives rise to a rich, four-state ferroelectric system under compressive strain. Figure~\ref{fig:strain}d shows the case of $\eta=-0.1$\%, where the four stable states form two pairs, distinguished by the buckling direction ($\Delta z_{\rm AA} > 0$ or $\Delta z_{\rm AA} < 0$). Crucially, an external electric field can only change the DW bending direction, thereby modulating the AB/BA domain ratio within a given buckling state; it cannot reverse the buckling direction of the AA regions (see the $\Delta z$-profile in the top panel of Fig.~\ref{fig:strain}d).
This constraint results in two distinct and asymmetric ferroelectric hysteresis loops, as confirmed by MD simulations (Fig.~\ref{fig:strain}c). The gray loop corresponds to the upward-buckled configuration ($\Delta z_{\rm AA} > 0$), where the electric field reversibly switches the system between a high-polarization state $+P_1$ and a low-polarization state $-P_2$ (top panel of Fig.~\ref{fig:strain}d). The yellow loop represents the downward-buckled configuration ($\Delta z_{\rm AA} < 0$), where the system switches between $-P_1$ and $+P_2$ (bottom panel of Fig.~\ref{fig:strain}d).
While an electric field alone cannot drive a transition between $+P_1$ and $-P_1$, 
we find that such a transition can be achieved mechanically. By locally applying force to the AA regions, the sign of $\Delta z_{\rm AA}$ can be reversed, enabling switching between two polar states of equal magnitude (left and right column of Fig.~\ref{fig:strain}d): $+P_1 \leftrightarrow -P_1$ and $+P_2 \leftrightarrow -P_2$. Thus, the combined use of electrical and mechanical stimuli allows full access to all four states, offering a complete control scheme for multi-level device applications.

\begin{figure}[t]
\includegraphics[width=0.95\textwidth]{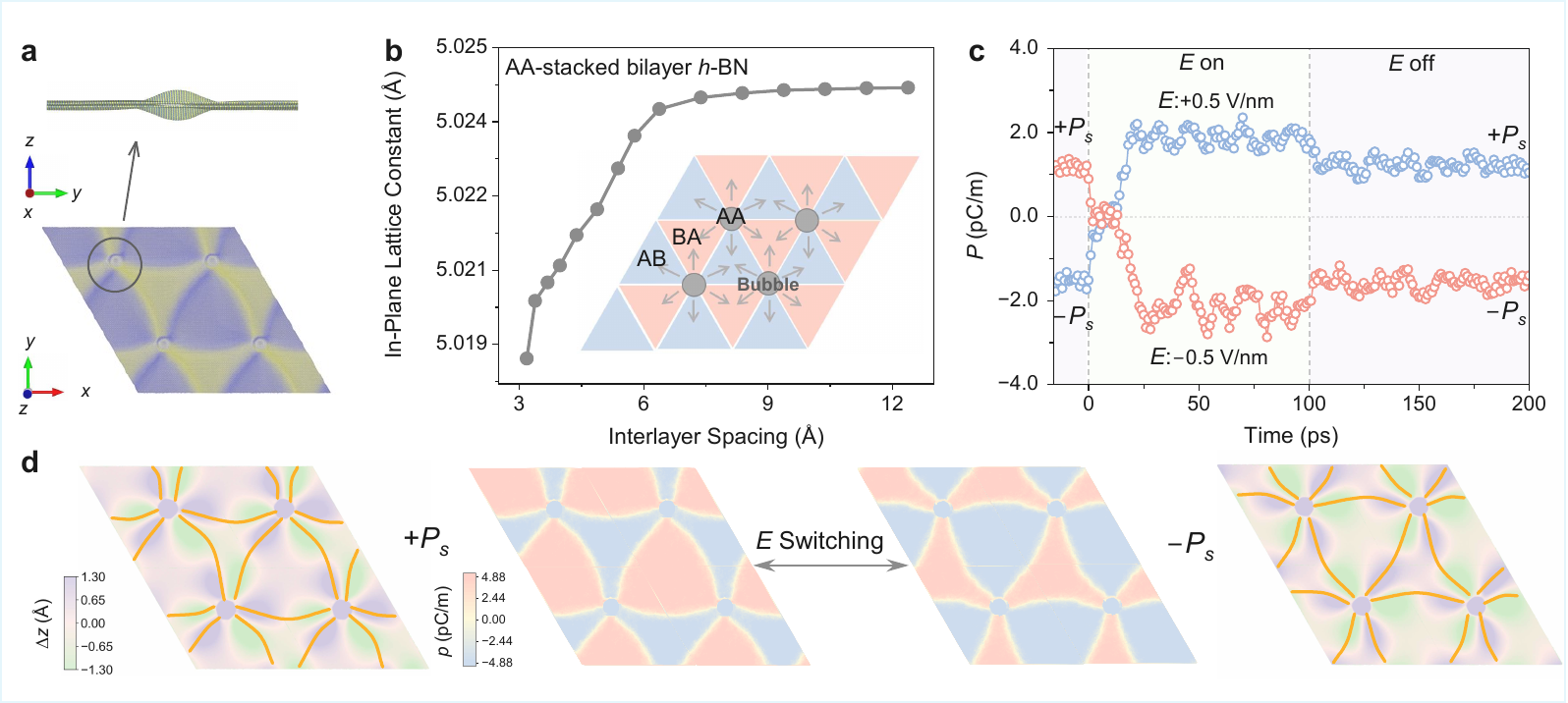}
\caption{\textbf{Nanobubble-induced global symmetry breaking and ferroelectric switching in moir\'e bilayer $h$-BN.} (\textbf{a}) 
Schematic of a moir\'e bilayer $h$-BN with a nanobubble pinned at each AA region. 
(\textbf{b}) Variation of the relaxed in-plane lattice constant with interlayer spacing in AA-stacked bilayer $h$-BN. The inset shows vertical expansion at the nanobubble center leads to local in-plane lattice expansion (gray arrows), which imposes compressive strain on adjacent AB/BA regions.
(\textbf{c}) Time evolution of the total polarization from MD simulations.
Starting from an equilibrium structure with nanobubbles pinned at AA sites, an external out-of-plane electric field ($E=\pm0.5$ V/nm) is applied and then removed. The system exhibits reversible switching between $+P_s$ and $-P_s$ states, with a stable remanent polarization after the field is turned off. (\textbf{d}) Local polarization ($\mu$) and buckling heights ($\Delta z$) for the two degenerated polar states that can be reversibly switched.}
\label{fig:bubble}
\end{figure}

Building on the concept of strain-induced polarization, we now explore nanobubbles as an experimentally relevant form of localized strain. In realistic samples, trace contaminants can become trapped between layers, forming stable blisters that act as local geometric defects~\cite{Blundo22p1525,wang19p116101}. We introduce nanobubbles into our MD simulations by applying opposite vertical forces to the same circular region of the top and bottom $h$-BN layers, creating a localized outward protrusion. The in-plane supercell dimensions are fixed to those of the fully relaxed, unstrained twisted bilayer. Our simulations at 300 K reveal a strong energetic preference for these nanobubbles to migrate and pin at the AA stacking sites (Fig.~\ref{fig:bubble}a), which coincide with the local C$_3$ rotational axes of the moir\'e pattern. This preferential pinning arises from interlayer energetics: while AB and BA stackings are energetically favorable, the AA stacking is less stable (Fig.~S4). A nanobubble preferentially stabilizes at an AA site due to the lower energy penalty compared to embedding it in AB or BA domains. This mechanism is consistent with the earlier observation that AA regions undergo the strongest outward buckling due to in-plane compressive strain.

We find that a nanobubble acts as a localized stressor, imposing an effective compressive strain on the surrounding AB and BA domains. This effect is elucidated by the model calculations in Fig.~\ref{fig:bubble}b, which show the relaxed in-plane lattice constant of an AA-stacked bilayer $h$-BN as a function of interlayer distance. As the layers are pushed apart, the in-plane lattice expands. Thus, a nanobubble at an AA site forces the local lattice to expand in-plane, effectively compressing adjacent AB/BA regions (see inset, Fig.\ref{fig:bubble}b).
We further construct a moir\'e supercell of 98284 atoms, placing a nanobubble at each AA site. Remarkably, the  nanobubble-patterned superlattice is ferroelectric, supporting two degenerate polar states, $+P_s$ and $-P_s$, of nearly equal magnitude, each featuring bent DWs and local C$_3$ rotational symmetry around every AA region (Fig.~\ref{fig:bubble}d). 
The corresponding real-space $\Delta z$-profiles reveal that these two states can be viewed as enantiomorphic partners, related by an out-of-plane mirror symmetry ($\sigma_h$) such that $\Delta z(\mathbf{r}) \rightarrow -\Delta z(\mathbf{r})$ under switching.
This bistability originates from the symmetric nature of the nanobubbles, which push both layers outward and generate local compressive strain while preserving local $\sigma_h$ and C$_3$ symmetry in $\Delta z$. 
As revealed by the time evolution of polarization in Fig.~\ref{fig:bubble}d from finite-field MD simulations, the $+P_s$ and $-P_s$ states can be reversibly switched, with a stable remanent polarization persisting after the field is removed.

Beyond inducing a switchable uniform global polarization, we discover that nanobubbles can be employed to create individually switchable ferroelectric bits. Each nanobubble pinned at a C$_3$ axis creates a localized imbalance in the surrounding AB/BA domain areas, giving rise to a net polarization that defines a ferroelectric bit within a single hexagonal moir\'e unit. This finding challenges the conventional view of moir\'e patterns as globally defined structures that are difficult to modify on a local scale. Our simulations further demonstrate that this local control is not only possible but also robust.

\begin{figure}[t]
\includegraphics[width=0.5\textwidth]{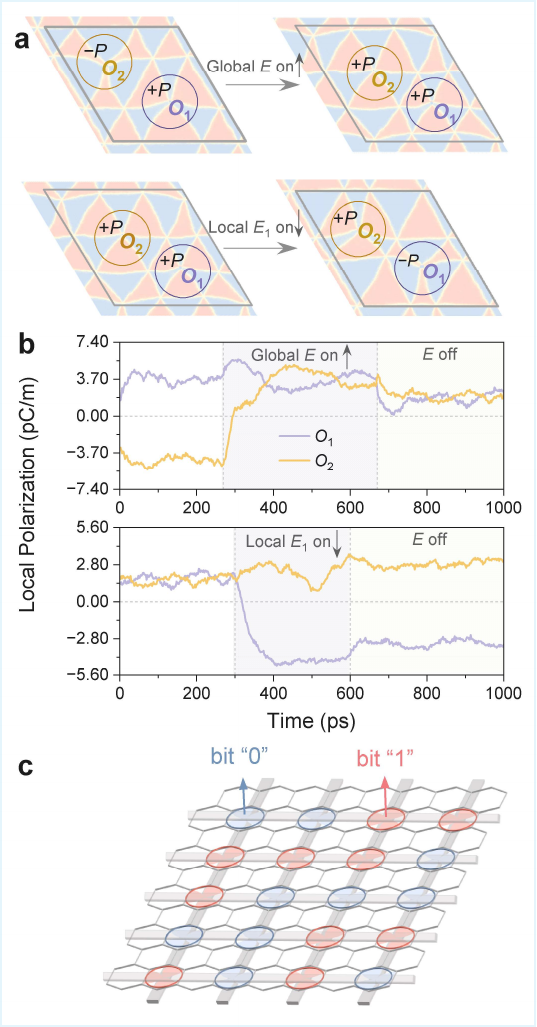}
\caption{\textbf{Individually addressable ferroelectric bits in moir\'e  bilayer $h$-BN}
(\textbf{a}) Domain patterns in supercells containing two nanobubbles centered at circular regions $O_1$ and $O_2$. The top panel illustrates the switching of $O_2$ from $-P$ to $+P$ under a global upward electric field. The bottom panel shows the switching of $O_1$ from $+P$ to $-P$ under a localized downward electric field $E_1$.
(\textbf{b}) Time evolution of local polarization in regions $O_1$ and $O_2$ corresponding to the processes shown in (\textbf{a}). 
(\textbf{c}) Conceptual device architecture resembling a crossbar array, where nanobubble-induced hexagonal moir\'e domains can be individually controlled to encode binary states ``0" and ``1".}
\label{fig:supercell}
\end{figure}

We simulate a 884556-atom supercell with in-plane dimensions of 60~\AA~$\times$~60~\AA~containing two nanobubbles. As shown in Fig.~\ref{fig:supercell}a, we initialize the system in a stable, mixed-state configuration where the two regions, labeled $O_1$ and $O_2$, exhibit opposite polarizations ($+P$ as ``1" and $-P$ as ``0", respectively). The top panel demonstrates that a global electric field can switch the $O_2$ bit from $-P$ to $+P$ while leaving the $O_1$ bit mostly undisturbed. The bottom panel shows that a downward local electric field ($E_1$), applied only to the $O_1$ region, reverses its polarization from $+P$ to $-P$ without affecting $O_2$. In both scenarios, the switched states are non-volatile, persisting long after the field is removed. The time evolution of the local polarization in regions $O_1$ and $O_2$ during the above processes, obtained from MD simulations, is shown in Fig.~\ref{fig:supercell}b. 
The spatially decoupled control arises from the highly localized nature of the nanobubble-induced curvature, which ensures that the state of one ferroelectric bit does not influence its neighbors.
These results confirm that the nanobubble-centered hexagonal moir\'e domain constitutes the smallest possible individually addressable ferroelectric bit. The independent addressability provides a blueprint for a crossbar memory architecture, as sketched in Fig.~\ref{fig:supercell}c, where each hexagonal moir\'e domain, with varying AB/BA domain ratios, can function as a discrete memory element representing bit ``0" or bit ``1". This establishes a physically grounded route toward ultra-high-density, reconfigurable ferroelectric devices based on engineered moir\'e patterns. Compared to previous proposals that rely on large-area moir\'e superlattices, our approach leverages much smaller, nanobubble-engineered hexagonal moir\'e domains, enabling finer spatial resolution and improved scalability.

In conclusion, we have unambiguously identified compressive-strain-induced out-of-plane rippling as the fundamental origin of moir\'e ferroelectricity, resolving the standing conflict between theoretical predictions and experimental observations. Our large-scale simulations demonstrate that this curvature field, rather than the widely assumed defect-pinning mechanism, breaks the global antiferroelectric symmetry by inducing domain wall bending to stabilize a net polarization. We further reveal that this mechanism can be harnessed at the nanoscale, where nanobubbles, potentially controllably engineered via trapped nanoparticles, act as localized strain sources. This creates an array of independently addressable ferroelectric bits, demonstrating that the global, topologically-resilient structure of a moir\'e superlattice can be locally manipulated. Our findings not only provide a robust blueprint for ultra-high-density memory but, more fundamentally, establish a geometry-driven paradigm for tuning the moir\'e potential itself, a crucial step toward designing and controlling the correlated and topological quantum phases hosted within these remarkable systems.\\

\begin{acknowledgments}
This work acknowledges the supports from National Natural Science Foundation of China (Grant No.~12304128) and Zhejiang Provincial Natural Science Foundation of China (LR25A040004). The computational resource is provided by Westlake HPC Center.
\end{acknowledgments}

\bibliography{SL}

\begin{thebibliography}{49}%
\makeatletter
\providecommand \@ifxundefined [1]{%
 \@ifx{#1\undefined}
}%
\providecommand \@ifnum [1]{%
 \ifnum #1\expandafter \@firstoftwo
 \else \expandafter \@secondoftwo
 \fi
}%
\providecommand \@ifx [1]{%
 \ifx #1\expandafter \@firstoftwo
 \else \expandafter \@secondoftwo
 \fi
}%
\providecommand \natexlab [1]{#1}%
\providecommand \enquote  [1]{``#1''}%
\providecommand \bibnamefont  [1]{#1}%
\providecommand \bibfnamefont [1]{#1}%
\providecommand \citenamefont [1]{#1}%
\providecommand \href@noop [0]{\@secondoftwo}%
\providecommand \href [0]{\begingroup \@sanitize@url \@href}%
\providecommand \@href[1]{\@@startlink{#1}\@@href}%
\providecommand \@@href[1]{\endgroup#1\@@endlink}%
\providecommand \@sanitize@url [0]{\catcode `\\12\catcode `\$12\catcode `\&12\catcode `\#12\catcode `\^12\catcode `\_12\catcode `\%12\relax}%
\providecommand \@@startlink[1]{}%
\providecommand \@@endlink[0]{}%
\providecommand \url  [0]{\begingroup\@sanitize@url \@url }%
\providecommand \@url [1]{\endgroup\@href {#1}{\urlprefix }}%
\providecommand \urlprefix  [0]{URL }%
\providecommand \Eprint [0]{\href }%
\providecommand \doibase [0]{https://doi.org/}%
\providecommand \selectlanguage [0]{\@gobble}%
\providecommand \bibinfo  [0]{\@secondoftwo}%
\providecommand \bibfield  [0]{\@secondoftwo}%
\providecommand \translation [1]{[#1]}%
\providecommand \BibitemOpen [0]{}%
\providecommand \bibitemStop [0]{}%
\providecommand \bibitemNoStop [0]{.\EOS\space}%
\providecommand \EOS [0]{\spacefactor3000\relax}%
\providecommand \BibitemShut  [1]{\csname bibitem#1\endcsname}%
\let\auto@bib@innerbib\@empty
\bibitem [{\citenamefont {Stern}\ \emph {et~al.}(2021)\citenamefont {Stern}, \citenamefont {Waschitz}, \citenamefont {Cao}, \citenamefont {Nevo}, \citenamefont {Watanabe}, \citenamefont {Taniguchi}, \citenamefont {Sela}, \citenamefont {Urbakh}, \citenamefont {Hod},\ and\ \citenamefont {Shalom}}]{ViznerStern21p1462}%
  \BibitemOpen
  \bibfield  {author} {\bibinfo {author} {\bibfnamefont {M.~V.}\ \bibnamefont {Stern}}, \bibinfo {author} {\bibfnamefont {Y.}~\bibnamefont {Waschitz}}, \bibinfo {author} {\bibfnamefont {W.}~\bibnamefont {Cao}}, \bibinfo {author} {\bibfnamefont {I.}~\bibnamefont {Nevo}}, \bibinfo {author} {\bibfnamefont {K.}~\bibnamefont {Watanabe}}, \bibinfo {author} {\bibfnamefont {T.}~\bibnamefont {Taniguchi}}, \bibinfo {author} {\bibfnamefont {E.}~\bibnamefont {Sela}}, \bibinfo {author} {\bibfnamefont {M.}~\bibnamefont {Urbakh}}, \bibinfo {author} {\bibfnamefont {O.}~\bibnamefont {Hod}},\ and\ \bibinfo {author} {\bibfnamefont {M.~B.}\ \bibnamefont {Shalom}},\ }\bibfield  {title} {\bibinfo {title} {Interfacial ferroelectricity by van der waals sliding},\ }\href {https://doi.org/10.1126/science.abe8177} {\bibfield  {journal} {\bibinfo  {journal} {Science}\ }\textbf {\bibinfo {volume} {372}},\ \bibinfo {pages} {1462} (\bibinfo {year} {2021})}\BibitemShut {NoStop}%
\bibitem [{\citenamefont {Yasuda}\ \emph {et~al.}(2021)\citenamefont {Yasuda}, \citenamefont {Wang}, \citenamefont {Watanabe}, \citenamefont {Taniguchi},\ and\ \citenamefont {Jarillo-Herrero}}]{Yasuda21p6549}%
  \BibitemOpen
  \bibfield  {author} {\bibinfo {author} {\bibfnamefont {K.}~\bibnamefont {Yasuda}}, \bibinfo {author} {\bibfnamefont {X.}~\bibnamefont {Wang}}, \bibinfo {author} {\bibfnamefont {K.}~\bibnamefont {Watanabe}}, \bibinfo {author} {\bibfnamefont {T.}~\bibnamefont {Taniguchi}},\ and\ \bibinfo {author} {\bibfnamefont {P.}~\bibnamefont {Jarillo-Herrero}},\ }\bibfield  {title} {\bibinfo {title} {Stacking-engineered ferroelectricity in bilayer boron nitride},\ }\href {https://doi.org/10.1126/science.abd3230} {\bibfield  {journal} {\bibinfo  {journal} {Science}\ }\textbf {\bibinfo {volume} {372}},\ \bibinfo {pages} {1458} (\bibinfo {year} {2021})}\BibitemShut {NoStop}%
\bibitem [{\citenamefont {Wu}\ and\ \citenamefont {Li}(2021)}]{Wu21pe2115703118}%
  \BibitemOpen
  \bibfield  {author} {\bibinfo {author} {\bibfnamefont {M.}~\bibnamefont {Wu}}\ and\ \bibinfo {author} {\bibfnamefont {J.}~\bibnamefont {Li}},\ }\bibfield  {title} {\bibinfo {title} {Sliding ferroelectricity in {2D} van der waals materials: Related physics and future opportunities},\ }\href {https://doi.org/10.1073/pnas.2115703118} {\bibfield  {journal} {\bibinfo  {journal} {Proc. Natl. Acad. Sci. U.S.A.}\ }\textbf {\bibinfo {volume} {118}},\ \bibinfo {pages} {e2115703118} (\bibinfo {year} {2021})}\BibitemShut {NoStop}%
\bibitem [{\citenamefont {Li}\ and\ \citenamefont {Wu}(2017)}]{Li17p6382}%
  \BibitemOpen
  \bibfield  {author} {\bibinfo {author} {\bibfnamefont {L.}~\bibnamefont {Li}}\ and\ \bibinfo {author} {\bibfnamefont {M.}~\bibnamefont {Wu}},\ }\bibfield  {title} {\bibinfo {title} {Binary compound bilayer and multilayer with vertical polarizations: Two-dimensional ferroelectrics, multiferroics, and nanogenerators},\ }\href {https://doi.org/10.1021/acsnano.7b02756} {\bibfield  {journal} {\bibinfo  {journal} {{ACS} Nano}\ }\textbf {\bibinfo {volume} {11}},\ \bibinfo {pages} {6382} (\bibinfo {year} {2017})}\BibitemShut {NoStop}%
\bibitem [{\citenamefont {Gao}\ and\ \citenamefont {Bellaiche}(2024)}]{Gao24p196801}%
  \BibitemOpen
  \bibfield  {author} {\bibinfo {author} {\bibfnamefont {L.}~\bibnamefont {Gao}}\ and\ \bibinfo {author} {\bibfnamefont {L.}~\bibnamefont {Bellaiche}},\ }\bibfield  {title} {\bibinfo {title} {Large photoinduced tuning of ferroelectricity in sliding ferroelectrics},\ }\href {https://doi.org/10.1103/PhysRevLett.133.196801} {\bibfield  {journal} {\bibinfo  {journal} {Phys. Rev. Lett.}\ }\textbf {\bibinfo {volume} {133}},\ \bibinfo {pages} {196801} (\bibinfo {year} {2024})}\BibitemShut {NoStop}%
\bibitem [{\citenamefont {Wang}\ \emph {et~al.}(2022)\citenamefont {Wang}, \citenamefont {Yasuda}, \citenamefont {Zhang}, \citenamefont {Liu}, \citenamefont {Watanabe}, \citenamefont {Taniguchi}, \citenamefont {Hone}, \citenamefont {Fu},\ and\ \citenamefont {Jarillo-Herrero}}]{Wang22p367}%
  \BibitemOpen
  \bibfield  {author} {\bibinfo {author} {\bibfnamefont {X.}~\bibnamefont {Wang}}, \bibinfo {author} {\bibfnamefont {K.}~\bibnamefont {Yasuda}}, \bibinfo {author} {\bibfnamefont {Y.}~\bibnamefont {Zhang}}, \bibinfo {author} {\bibfnamefont {S.}~\bibnamefont {Liu}}, \bibinfo {author} {\bibfnamefont {K.}~\bibnamefont {Watanabe}}, \bibinfo {author} {\bibfnamefont {T.}~\bibnamefont {Taniguchi}}, \bibinfo {author} {\bibfnamefont {J.}~\bibnamefont {Hone}}, \bibinfo {author} {\bibfnamefont {L.}~\bibnamefont {Fu}},\ and\ \bibinfo {author} {\bibfnamefont {P.}~\bibnamefont {Jarillo-Herrero}},\ }\bibfield  {title} {\bibinfo {title} {Interfacial ferroelectricity in rhombohedral-stacked bilayer transition metal dichalcogenides},\ }\href {https://doi.org/10.1038/s41565-021-01059-z} {\bibfield  {journal} {\bibinfo  {journal} {Nat. Nanotechnol.}\ }\textbf {\bibinfo {volume} {17}},\ \bibinfo {pages} {367} (\bibinfo {year} {2022})}\BibitemShut {NoStop}%
\bibitem [{\citenamefont {Weston}\ \emph {et~al.}(2022)\citenamefont {Weston}, \citenamefont {Castanon}, \citenamefont {Enaldiev}, \citenamefont {Ferreira}, \citenamefont {Bhattacharjee}, \citenamefont {Xu}, \citenamefont {Corte-León}, \citenamefont {Wu}, \citenamefont {Clark}, \citenamefont {Summerfield}, \citenamefont {Hashimoto}, \citenamefont {Gao}, \citenamefont {Wang}, \citenamefont {Hamer}, \citenamefont {Read}, \citenamefont {Fumagalli}, \citenamefont {Kretinin}, \citenamefont {Haigh}, \citenamefont {Kazakova}, \citenamefont {Geim}, \citenamefont {Fal’ko},\ and\ \citenamefont {Gorbachev}}]{Weston22p390}%
  \BibitemOpen
  \bibfield  {author} {\bibinfo {author} {\bibfnamefont {A.}~\bibnamefont {Weston}}, \bibinfo {author} {\bibfnamefont {E.~G.}\ \bibnamefont {Castanon}}, \bibinfo {author} {\bibfnamefont {V.}~\bibnamefont {Enaldiev}}, \bibinfo {author} {\bibfnamefont {F.}~\bibnamefont {Ferreira}}, \bibinfo {author} {\bibfnamefont {S.}~\bibnamefont {Bhattacharjee}}, \bibinfo {author} {\bibfnamefont {S.}~\bibnamefont {Xu}}, \bibinfo {author} {\bibfnamefont {H.}~\bibnamefont {Corte-León}}, \bibinfo {author} {\bibfnamefont {Z.}~\bibnamefont {Wu}}, \bibinfo {author} {\bibfnamefont {N.}~\bibnamefont {Clark}}, \bibinfo {author} {\bibfnamefont {A.}~\bibnamefont {Summerfield}}, \bibinfo {author} {\bibfnamefont {T.}~\bibnamefont {Hashimoto}}, \bibinfo {author} {\bibfnamefont {Y.}~\bibnamefont {Gao}}, \bibinfo {author} {\bibfnamefont {W.}~\bibnamefont {Wang}}, \bibinfo {author} {\bibfnamefont {M.}~\bibnamefont {Hamer}}, \bibinfo {author} {\bibfnamefont {H.}~\bibnamefont {Read}}, \bibinfo {author} {\bibfnamefont {L.}~\bibnamefont
  {Fumagalli}}, \bibinfo {author} {\bibfnamefont {A.~V.}\ \bibnamefont {Kretinin}}, \bibinfo {author} {\bibfnamefont {S.~J.}\ \bibnamefont {Haigh}}, \bibinfo {author} {\bibfnamefont {O.}~\bibnamefont {Kazakova}}, \bibinfo {author} {\bibfnamefont {A.~K.}\ \bibnamefont {Geim}}, \bibinfo {author} {\bibfnamefont {V.~I.}\ \bibnamefont {Fal’ko}},\ and\ \bibinfo {author} {\bibfnamefont {R.}~\bibnamefont {Gorbachev}},\ }\bibfield  {title} {\bibinfo {title} {Interfacial ferroelectricity in marginally twisted {2D} semiconductors},\ }\href {https://www.nature.com/articles/s41565-022-01072-w} {\bibfield  {journal} {\bibinfo  {journal} {Nat. Nanotechnol.}\ }\textbf {\bibinfo {volume} {17}},\ \bibinfo {pages} {390} (\bibinfo {year} {2022})}\BibitemShut {NoStop}%
\bibitem [{\citenamefont {Van~Winkle}\ \emph {et~al.}(2024)\citenamefont {Van~Winkle}, \citenamefont {Dowlatshahi}, \citenamefont {Khaloo}, \citenamefont {Iyer}, \citenamefont {Craig}, \citenamefont {Dhall}, \citenamefont {Taniguchi}, \citenamefont {Watanabe},\ and\ \citenamefont {Bediako}}]{VanWinkle24p751}%
  \BibitemOpen
  \bibfield  {author} {\bibinfo {author} {\bibfnamefont {M.}~\bibnamefont {Van~Winkle}}, \bibinfo {author} {\bibfnamefont {N.}~\bibnamefont {Dowlatshahi}}, \bibinfo {author} {\bibfnamefont {N.}~\bibnamefont {Khaloo}}, \bibinfo {author} {\bibfnamefont {M.}~\bibnamefont {Iyer}}, \bibinfo {author} {\bibfnamefont {I.~M.}\ \bibnamefont {Craig}}, \bibinfo {author} {\bibfnamefont {R.}~\bibnamefont {Dhall}}, \bibinfo {author} {\bibfnamefont {T.}~\bibnamefont {Taniguchi}}, \bibinfo {author} {\bibfnamefont {K.}~\bibnamefont {Watanabe}},\ and\ \bibinfo {author} {\bibfnamefont {D.~K.}\ \bibnamefont {Bediako}},\ }\bibfield  {title} {\bibinfo {title} {Engineering interfacial polarization switching in van der {Waals} multilayers},\ }\href {https://doi.org/10.1038/s41565-024-01642-0} {\bibfield  {journal} {\bibinfo  {journal} {Nat. Nanotechnol.}\ }\textbf {\bibinfo {volume} {19}},\ \bibinfo {pages} {751} (\bibinfo {year} {2024})}\BibitemShut {NoStop}%
\bibitem [{\citenamefont {Guo}\ \emph {et~al.}(2025)\citenamefont {Guo}, \citenamefont {Yan}, \citenamefont {He}, \citenamefont {Lv}, \citenamefont {Watanabe}, \citenamefont {Taniguchi}, \citenamefont {Ren},\ and\ \citenamefont {He}}]{guo25p11543}%
  \BibitemOpen
  \bibfield  {author} {\bibinfo {author} {\bibfnamefont {Z.-H.}\ \bibnamefont {Guo}}, \bibinfo {author} {\bibfnamefont {C.}~\bibnamefont {Yan}}, \bibinfo {author} {\bibfnamefont {J.-Q.}\ \bibnamefont {He}}, \bibinfo {author} {\bibfnamefont {K.}~\bibnamefont {Lv}}, \bibinfo {author} {\bibfnamefont {K.}~\bibnamefont {Watanabe}}, \bibinfo {author} {\bibfnamefont {T.}~\bibnamefont {Taniguchi}}, \bibinfo {author} {\bibfnamefont {Y.-N.}\ \bibnamefont {Ren}},\ and\ \bibinfo {author} {\bibfnamefont {L.}~\bibnamefont {He}},\ }\bibfield  {title} {\bibinfo {title} {Manipulating superlattice potentials and quantum confinement in graphene via moir{\'e} ferroelectricity},\ }\href {https://doi.org/10.1021/acs.nanolett.5c01976} {\bibfield  {journal} {\bibinfo  {journal} {Nano Lett.}\ }\textbf {\bibinfo {volume} {25}},\ \bibinfo {pages} {11543–11548} (\bibinfo {year} {2025})}\BibitemShut {NoStop}%
\bibitem [{\citenamefont {Du}\ \emph {et~al.}(2024)\citenamefont {Du}, \citenamefont {Xiao}, \citenamefont {Zhang}, \citenamefont {Cai}, \citenamefont {Jiang}, \citenamefont {Lian}, \citenamefont {Watanabe}, \citenamefont {Taniguchi}, \citenamefont {Wang},\ and\ \citenamefont {Yu}}]{du24p025015}%
  \BibitemOpen
  \bibfield  {author} {\bibinfo {author} {\bibfnamefont {R.}~\bibnamefont {Du}}, \bibinfo {author} {\bibfnamefont {J.}~\bibnamefont {Xiao}}, \bibinfo {author} {\bibfnamefont {D.}~\bibnamefont {Zhang}}, \bibinfo {author} {\bibfnamefont {X.}~\bibnamefont {Cai}}, \bibinfo {author} {\bibfnamefont {S.}~\bibnamefont {Jiang}}, \bibinfo {author} {\bibfnamefont {F.}~\bibnamefont {Lian}}, \bibinfo {author} {\bibfnamefont {K.}~\bibnamefont {Watanabe}}, \bibinfo {author} {\bibfnamefont {T.}~\bibnamefont {Taniguchi}}, \bibinfo {author} {\bibfnamefont {L.}~\bibnamefont {Wang}},\ and\ \bibinfo {author} {\bibfnamefont {G.}~\bibnamefont {Yu}},\ }\bibfield  {title} {\bibinfo {title} {Ferroelectricity in twisted double bilayer graphene},\ }\href {https://doi.org/10.1088/2053-1583/ad2107} {\bibfield  {journal} {\bibinfo  {journal} {2D Mater.}\ }\textbf {\bibinfo {volume} {11}},\ \bibinfo {pages} {025015} (\bibinfo {year} {2024})}\BibitemShut {NoStop}%
\bibitem [{\citenamefont {Zheng}\ \emph {et~al.}(2020)\citenamefont {Zheng}, \citenamefont {Ma}, \citenamefont {Bi}, \citenamefont {de~la Barrera}, \citenamefont {Liu}, \citenamefont {Mao}, \citenamefont {Zhang}, \citenamefont {Kiper}, \citenamefont {Watanabe}, \citenamefont {Taniguchi}, \citenamefont {Kong}, \citenamefont {Tisdale}, \citenamefont {Ashoori}, \citenamefont {Gedik}, \citenamefont {Fu}, \citenamefont {Xu},\ and\ \citenamefont {Jarillo-Herrero}}]{Zheng20p71}%
  \BibitemOpen
  \bibfield  {author} {\bibinfo {author} {\bibfnamefont {Z.}~\bibnamefont {Zheng}}, \bibinfo {author} {\bibfnamefont {Q.}~\bibnamefont {Ma}}, \bibinfo {author} {\bibfnamefont {Z.}~\bibnamefont {Bi}}, \bibinfo {author} {\bibfnamefont {S.}~\bibnamefont {de~la Barrera}}, \bibinfo {author} {\bibfnamefont {M.-H.}\ \bibnamefont {Liu}}, \bibinfo {author} {\bibfnamefont {N.}~\bibnamefont {Mao}}, \bibinfo {author} {\bibfnamefont {Y.}~\bibnamefont {Zhang}}, \bibinfo {author} {\bibfnamefont {N.}~\bibnamefont {Kiper}}, \bibinfo {author} {\bibfnamefont {K.}~\bibnamefont {Watanabe}}, \bibinfo {author} {\bibfnamefont {T.}~\bibnamefont {Taniguchi}}, \bibinfo {author} {\bibfnamefont {J.}~\bibnamefont {Kong}}, \bibinfo {author} {\bibfnamefont {W.~A.}\ \bibnamefont {Tisdale}}, \bibinfo {author} {\bibfnamefont {R.}~\bibnamefont {Ashoori}}, \bibinfo {author} {\bibfnamefont {N.}~\bibnamefont {Gedik}}, \bibinfo {author} {\bibfnamefont {L.}~\bibnamefont {Fu}}, \bibinfo {author} {\bibfnamefont {S.-Y.}\ \bibnamefont {Xu}},\ and\
  \bibinfo {author} {\bibfnamefont {P.}~\bibnamefont {Jarillo-Herrero}},\ }\bibfield  {title} {\bibinfo {title} {Unconventional ferroelectricity in moir{\'{e}} heterostructures},\ }\href {https://doi.org/10.1038/s41586-020-2970-9} {\bibfield  {journal} {\bibinfo  {journal} {Nature}\ }\textbf {\bibinfo {volume} {588}},\ \bibinfo {pages} {71} (\bibinfo {year} {2020})}\BibitemShut {NoStop}%
\bibitem [{\citenamefont {Niu}\ \emph {et~al.}(2022)\citenamefont {Niu}, \citenamefont {Li}, \citenamefont {Han}, \citenamefont {Qu}, \citenamefont {Ding}, \citenamefont {Wang}, \citenamefont {Liu}, \citenamefont {Liu}, \citenamefont {Han}, \citenamefont {Watanabe} \emph {et~al.}}]{niu22p6241}%
  \BibitemOpen
  \bibfield  {author} {\bibinfo {author} {\bibfnamefont {R.}~\bibnamefont {Niu}}, \bibinfo {author} {\bibfnamefont {Z.}~\bibnamefont {Li}}, \bibinfo {author} {\bibfnamefont {X.}~\bibnamefont {Han}}, \bibinfo {author} {\bibfnamefont {Z.}~\bibnamefont {Qu}}, \bibinfo {author} {\bibfnamefont {D.}~\bibnamefont {Ding}}, \bibinfo {author} {\bibfnamefont {Z.}~\bibnamefont {Wang}}, \bibinfo {author} {\bibfnamefont {Q.}~\bibnamefont {Liu}}, \bibinfo {author} {\bibfnamefont {T.}~\bibnamefont {Liu}}, \bibinfo {author} {\bibfnamefont {C.}~\bibnamefont {Han}}, \bibinfo {author} {\bibfnamefont {K.}~\bibnamefont {Watanabe}}, \emph {et~al.},\ }\bibfield  {title} {\bibinfo {title} {Giant ferroelectric polarization in a bilayer graphene heterostructure},\ }\href {https://doi.org/10.1038/s41467-022-34104-z} {\bibfield  {journal} {\bibinfo  {journal} {Nat. Commun.}\ }\textbf {\bibinfo {volume} {13}},\ \bibinfo {pages} {6241} (\bibinfo {year} {2022})}\BibitemShut {NoStop}%
\bibitem [{\citenamefont {Cao}\ \emph {et~al.}(2018{\natexlab{a}})\citenamefont {Cao}, \citenamefont {Fatemi}, \citenamefont {Fang}, \citenamefont {Watanabe}, \citenamefont {Taniguchi}, \citenamefont {Kaxiras},\ and\ \citenamefont {Jarillo-Herrero}}]{Cao18p43}%
  \BibitemOpen
  \bibfield  {author} {\bibinfo {author} {\bibfnamefont {Y.}~\bibnamefont {Cao}}, \bibinfo {author} {\bibfnamefont {V.}~\bibnamefont {Fatemi}}, \bibinfo {author} {\bibfnamefont {S.}~\bibnamefont {Fang}}, \bibinfo {author} {\bibfnamefont {K.}~\bibnamefont {Watanabe}}, \bibinfo {author} {\bibfnamefont {T.}~\bibnamefont {Taniguchi}}, \bibinfo {author} {\bibfnamefont {E.}~\bibnamefont {Kaxiras}},\ and\ \bibinfo {author} {\bibfnamefont {P.}~\bibnamefont {Jarillo-Herrero}},\ }\bibfield  {title} {\bibinfo {title} {Unconventional superconductivity in magic-angle graphene superlattices},\ }\href {https://doi.org/10.1038/nature26160} {\bibfield  {journal} {\bibinfo  {journal} {Nature}\ }\textbf {\bibinfo {volume} {556}},\ \bibinfo {pages} {43} (\bibinfo {year} {2018}{\natexlab{a}})}\BibitemShut {NoStop}%
\bibitem [{\citenamefont {Can}\ \emph {et~al.}(2021)\citenamefont {Can}, \citenamefont {Tummuru}, \citenamefont {Day}, \citenamefont {Elfimov}, \citenamefont {Damascelli},\ and\ \citenamefont {Franz}}]{can21p519}%
  \BibitemOpen
  \bibfield  {author} {\bibinfo {author} {\bibfnamefont {O.}~\bibnamefont {Can}}, \bibinfo {author} {\bibfnamefont {T.}~\bibnamefont {Tummuru}}, \bibinfo {author} {\bibfnamefont {R.~P.}\ \bibnamefont {Day}}, \bibinfo {author} {\bibfnamefont {I.}~\bibnamefont {Elfimov}}, \bibinfo {author} {\bibfnamefont {A.}~\bibnamefont {Damascelli}},\ and\ \bibinfo {author} {\bibfnamefont {M.}~\bibnamefont {Franz}},\ }\bibfield  {title} {\bibinfo {title} {High-temperature topological superconductivity in twisted double-layer copper oxides},\ }\href {https://doi.org/10.1038/s41567-020-01142-7} {\bibfield  {journal} {\bibinfo  {journal} {Nat. Phys.}\ }\textbf {\bibinfo {volume} {17}},\ \bibinfo {pages} {519} (\bibinfo {year} {2021})}\BibitemShut {NoStop}%
\bibitem [{\citenamefont {Song}\ \emph {et~al.}(2019)\citenamefont {Song}, \citenamefont {Wang}, \citenamefont {Shi}, \citenamefont {Li}, \citenamefont {Fang},\ and\ \citenamefont {Bernevig}}]{song19p123}%
  \BibitemOpen
  \bibfield  {author} {\bibinfo {author} {\bibfnamefont {Z.}~\bibnamefont {Song}}, \bibinfo {author} {\bibfnamefont {Z.}~\bibnamefont {Wang}}, \bibinfo {author} {\bibfnamefont {W.}~\bibnamefont {Shi}}, \bibinfo {author} {\bibfnamefont {G.}~\bibnamefont {Li}}, \bibinfo {author} {\bibfnamefont {C.}~\bibnamefont {Fang}},\ and\ \bibinfo {author} {\bibfnamefont {B.~A.}\ \bibnamefont {Bernevig}},\ }\bibfield  {title} {\bibinfo {title} {All magic angles in twisted bilayer graphene are topological},\ }\href {https://doi.org/10.1103/PhysRevLett.123.036401} {\bibfield  {journal} {\bibinfo  {journal} {Phys. Rev. Lett.}\ }\textbf {\bibinfo {volume} {123}},\ \bibinfo {pages} {036401} (\bibinfo {year} {2019})}\BibitemShut {NoStop}%
\bibitem [{\citenamefont {Lian}\ \emph {et~al.}(2020)\citenamefont {Lian}, \citenamefont {Liu}, \citenamefont {Zhang},\ and\ \citenamefont {Wang}}]{Lian20p12}%
  \BibitemOpen
  \bibfield  {author} {\bibinfo {author} {\bibfnamefont {B.}~\bibnamefont {Lian}}, \bibinfo {author} {\bibfnamefont {Z.}~\bibnamefont {Liu}}, \bibinfo {author} {\bibfnamefont {Y.}~\bibnamefont {Zhang}},\ and\ \bibinfo {author} {\bibfnamefont {J.}~\bibnamefont {Wang}},\ }\bibfield  {title} {\bibinfo {title} {Flat chern band from twisted bilayer {MnBi$_2$Te$_4$}},\ }\href {https://doi.org/10.1103/PhysRevLett.124.126402} {\bibfield  {journal} {\bibinfo  {journal} {Phys. Rev. Lett.}\ }\textbf {\bibinfo {volume} {124}},\ \bibinfo {pages} {126402} (\bibinfo {year} {2020})}\BibitemShut {NoStop}%
\bibitem [{\citenamefont {Serlin}\ \emph {et~al.}(2020)\citenamefont {Serlin}, \citenamefont {Tschirhart}, \citenamefont {Polshyn}, \citenamefont {Zhang}, \citenamefont {Zhu}, \citenamefont {Watanabe}, \citenamefont {Taniguchi}, \citenamefont {Balents},\ and\ \citenamefont {Young}}]{serlin20p367}%
  \BibitemOpen
  \bibfield  {author} {\bibinfo {author} {\bibfnamefont {M.}~\bibnamefont {Serlin}}, \bibinfo {author} {\bibfnamefont {C.}~\bibnamefont {Tschirhart}}, \bibinfo {author} {\bibfnamefont {H.}~\bibnamefont {Polshyn}}, \bibinfo {author} {\bibfnamefont {Y.}~\bibnamefont {Zhang}}, \bibinfo {author} {\bibfnamefont {J.}~\bibnamefont {Zhu}}, \bibinfo {author} {\bibfnamefont {K.}~\bibnamefont {Watanabe}}, \bibinfo {author} {\bibfnamefont {T.}~\bibnamefont {Taniguchi}}, \bibinfo {author} {\bibfnamefont {L.}~\bibnamefont {Balents}},\ and\ \bibinfo {author} {\bibfnamefont {A.}~\bibnamefont {Young}},\ }\bibfield  {title} {\bibinfo {title} {Intrinsic quantized anomalous hall effect in a moir{\'e} heterostructure},\ }\href {https://doi.org/10.1126/science.aay5533} {\bibfield  {journal} {\bibinfo  {journal} {Science}\ }\textbf {\bibinfo {volume} {367}},\ \bibinfo {pages} {900} (\bibinfo {year} {2020})}\BibitemShut {NoStop}%
\bibitem [{\citenamefont {Wu}\ \emph {et~al.}(2023)\citenamefont {Wu}, \citenamefont {Li}, \citenamefont {Xu}, \citenamefont {Liu}, \citenamefont {Yuan}, \citenamefont {Zhao}, \citenamefont {Huang}, \citenamefont {Zan}, \citenamefont {Watanabe}, \citenamefont {Taniguchi}, \citenamefont {Shi}, \citenamefont {Xian}, \citenamefont {Yang}, \citenamefont {Du},\ and\ \citenamefont {Zhang}}]{Wu23p4}%
  \BibitemOpen
  \bibfield  {author} {\bibinfo {author} {\bibfnamefont {F.}~\bibnamefont {Wu}}, \bibinfo {author} {\bibfnamefont {L.}~\bibnamefont {Li}}, \bibinfo {author} {\bibfnamefont {Q.}~\bibnamefont {Xu}}, \bibinfo {author} {\bibfnamefont {L.}~\bibnamefont {Liu}}, \bibinfo {author} {\bibfnamefont {Y.}~\bibnamefont {Yuan}}, \bibinfo {author} {\bibfnamefont {J.}~\bibnamefont {Zhao}}, \bibinfo {author} {\bibfnamefont {Z.}~\bibnamefont {Huang}}, \bibinfo {author} {\bibfnamefont {X.}~\bibnamefont {Zan}}, \bibinfo {author} {\bibfnamefont {K.}~\bibnamefont {Watanabe}}, \bibinfo {author} {\bibfnamefont {T.}~\bibnamefont {Taniguchi}}, \bibinfo {author} {\bibfnamefont {D.}~\bibnamefont {Shi}}, \bibinfo {author} {\bibfnamefont {L.}~\bibnamefont {Xian}}, \bibinfo {author} {\bibfnamefont {W.}~\bibnamefont {Yang}}, \bibinfo {author} {\bibfnamefont {L.}~\bibnamefont {Du}},\ and\ \bibinfo {author} {\bibfnamefont {G.}~\bibnamefont {Zhang}},\ }\bibfield  {title} {\bibinfo {title} {Coupled ferroelectricity and correlated states in a
  twisted quadrilayer {MoS$_2$} moiré superlattice},\ }\href {https://doi.org/10.1088/0256-307X/40/4/047303} {\bibfield  {journal} {\bibinfo  {journal} {Chin. Phys. Lett.}\ }\textbf {\bibinfo {volume} {40}} (\bibinfo {year} {2023})}\BibitemShut {NoStop}%
\bibitem [{\citenamefont {Kang}\ \emph {et~al.}(2023)\citenamefont {Kang}, \citenamefont {Zhao}, \citenamefont {Zeng}, \citenamefont {Watanabe}, \citenamefont {Taniguchi}, \citenamefont {Shan},\ and\ \citenamefont {Mak}}]{kang23p861}%
  \BibitemOpen
  \bibfield  {author} {\bibinfo {author} {\bibfnamefont {K.}~\bibnamefont {Kang}}, \bibinfo {author} {\bibfnamefont {W.}~\bibnamefont {Zhao}}, \bibinfo {author} {\bibfnamefont {Y.}~\bibnamefont {Zeng}}, \bibinfo {author} {\bibfnamefont {K.}~\bibnamefont {Watanabe}}, \bibinfo {author} {\bibfnamefont {T.}~\bibnamefont {Taniguchi}}, \bibinfo {author} {\bibfnamefont {J.}~\bibnamefont {Shan}},\ and\ \bibinfo {author} {\bibfnamefont {K.~F.}\ \bibnamefont {Mak}},\ }\bibfield  {title} {\bibinfo {title} {Switchable moir{\'e} potentials in ferroelectric {WTe$_2$}/{WSe$_2$} superlattices},\ }\href {https://doi.org/10.1038/s41565-023-01376-5} {\bibfield  {journal} {\bibinfo  {journal} {Nat. Nanotechnol.}\ }\textbf {\bibinfo {volume} {18}},\ \bibinfo {pages} {861} (\bibinfo {year} {2023})}\BibitemShut {NoStop}%
\bibitem [{\citenamefont {Cao}\ \emph {et~al.}(2018{\natexlab{b}})\citenamefont {Cao}, \citenamefont {Fatemi}, \citenamefont {Demir}, \citenamefont {Fang}, \citenamefont {Tomarken}, \citenamefont {Luo}, \citenamefont {Sanchez-Yamagishi}, \citenamefont {Watanabe}, \citenamefont {Taniguchi}, \citenamefont {Kaxiras}, \citenamefont {Ashoori},\ and\ \citenamefont {Jarillo-Herrero}}]{Cao18p80}%
  \BibitemOpen
  \bibfield  {author} {\bibinfo {author} {\bibfnamefont {Y.}~\bibnamefont {Cao}}, \bibinfo {author} {\bibfnamefont {V.}~\bibnamefont {Fatemi}}, \bibinfo {author} {\bibfnamefont {A.}~\bibnamefont {Demir}}, \bibinfo {author} {\bibfnamefont {S.}~\bibnamefont {Fang}}, \bibinfo {author} {\bibfnamefont {S.~L.}\ \bibnamefont {Tomarken}}, \bibinfo {author} {\bibfnamefont {J.~Y.}\ \bibnamefont {Luo}}, \bibinfo {author} {\bibfnamefont {J.~D.}\ \bibnamefont {Sanchez-Yamagishi}}, \bibinfo {author} {\bibfnamefont {K.}~\bibnamefont {Watanabe}}, \bibinfo {author} {\bibfnamefont {T.}~\bibnamefont {Taniguchi}}, \bibinfo {author} {\bibfnamefont {E.}~\bibnamefont {Kaxiras}}, \bibinfo {author} {\bibfnamefont {R.~C.}\ \bibnamefont {Ashoori}},\ and\ \bibinfo {author} {\bibfnamefont {P.}~\bibnamefont {Jarillo-Herrero}},\ }\bibfield  {title} {\bibinfo {title} {Correlated insulator behaviour at half-filling in magic-angle graphene superlattices},\ }\href {https://doi.org/10.1038/nature26154} {\bibfield  {journal} {\bibinfo  {journal}
  {Nature}\ }\textbf {\bibinfo {volume} {556}},\ \bibinfo {pages} {80} (\bibinfo {year} {2018}{\natexlab{b}})}\BibitemShut {NoStop}%
\bibitem [{\citenamefont {Weston}\ \emph {et~al.}(2020)\citenamefont {Weston}, \citenamefont {Zou}, \citenamefont {Enaldiev}, \citenamefont {Summerfield}, \citenamefont {Clark}, \citenamefont {Z{\'o}lyomi}, \citenamefont {Graham}, \citenamefont {Yelgel}, \citenamefont {Magorrian}, \citenamefont {Zhou} \emph {et~al.}}]{Weston20p592}%
  \BibitemOpen
  \bibfield  {author} {\bibinfo {author} {\bibfnamefont {A.}~\bibnamefont {Weston}}, \bibinfo {author} {\bibfnamefont {Y.}~\bibnamefont {Zou}}, \bibinfo {author} {\bibfnamefont {V.}~\bibnamefont {Enaldiev}}, \bibinfo {author} {\bibfnamefont {A.}~\bibnamefont {Summerfield}}, \bibinfo {author} {\bibfnamefont {N.}~\bibnamefont {Clark}}, \bibinfo {author} {\bibfnamefont {V.}~\bibnamefont {Z{\'o}lyomi}}, \bibinfo {author} {\bibfnamefont {A.}~\bibnamefont {Graham}}, \bibinfo {author} {\bibfnamefont {C.}~\bibnamefont {Yelgel}}, \bibinfo {author} {\bibfnamefont {S.}~\bibnamefont {Magorrian}}, \bibinfo {author} {\bibfnamefont {M.}~\bibnamefont {Zhou}}, \emph {et~al.},\ }\bibfield  {title} {\bibinfo {title} {Atomic reconstruction in twisted bilayers of transition metal dichalcogenides},\ }\href {https://doi.org/10.1038/s41565-020-0682-9} {\bibfield  {journal} {\bibinfo  {journal} {Nat. Nanotechnol.}\ }\textbf {\bibinfo {volume} {15}},\ \bibinfo {pages} {592} (\bibinfo {year} {2020})}\BibitemShut {NoStop}%
\bibitem [{\citenamefont {Bennett}(2022)}]{Bennett22p235445}%
  \BibitemOpen
  \bibfield  {author} {\bibinfo {author} {\bibfnamefont {D.}~\bibnamefont {Bennett}},\ }\bibfield  {title} {\bibinfo {title} {Theory of polar domains in moir\'e heterostructures},\ }\href {https://doi.org/10.1103/PhysRevB.105.235445} {\bibfield  {journal} {\bibinfo  {journal} {Phys. Rev. B}\ }\textbf {\bibinfo {volume} {105}},\ \bibinfo {pages} {235445} (\bibinfo {year} {2022})}\BibitemShut {NoStop}%
\bibitem [{\citenamefont {Moore}\ \emph {et~al.}(2021)\citenamefont {Moore}, \citenamefont {Ciccarino}, \citenamefont {Halbertal}, \citenamefont {McGilly}, \citenamefont {Finney}, \citenamefont {Yao}, \citenamefont {Shao}, \citenamefont {Ni}, \citenamefont {Sternbach}, \citenamefont {Telford} \emph {et~al.}}]{moore21p5741}%
  \BibitemOpen
  \bibfield  {author} {\bibinfo {author} {\bibfnamefont {S.}~\bibnamefont {Moore}}, \bibinfo {author} {\bibfnamefont {C.}~\bibnamefont {Ciccarino}}, \bibinfo {author} {\bibfnamefont {D.}~\bibnamefont {Halbertal}}, \bibinfo {author} {\bibfnamefont {L.}~\bibnamefont {McGilly}}, \bibinfo {author} {\bibfnamefont {N.}~\bibnamefont {Finney}}, \bibinfo {author} {\bibfnamefont {K.}~\bibnamefont {Yao}}, \bibinfo {author} {\bibfnamefont {Y.}~\bibnamefont {Shao}}, \bibinfo {author} {\bibfnamefont {G.}~\bibnamefont {Ni}}, \bibinfo {author} {\bibfnamefont {A.}~\bibnamefont {Sternbach}}, \bibinfo {author} {\bibfnamefont {E.}~\bibnamefont {Telford}}, \emph {et~al.},\ }\bibfield  {title} {\bibinfo {title} {Nanoscale lattice dynamics in hexagonal boron nitride moir{\'e} superlattices},\ }\href {https://doi.org/10.1038/s41467-021-26072-7} {\bibfield  {journal} {\bibinfo  {journal} {Nat. Commun.}\ }\textbf {\bibinfo {volume} {12}},\ \bibinfo {pages} {5741} (\bibinfo {year} {2021})}\BibitemShut {NoStop}%
\bibitem [{\citenamefont {Dang}\ \emph {et~al.}(2025)\citenamefont {Dang}, \citenamefont {Le},\ and\ \citenamefont {Woods}}]{Dang25p085102}%
  \BibitemOpen
  \bibfield  {author} {\bibinfo {author} {\bibfnamefont {D.~T.-X.}\ \bibnamefont {Dang}}, \bibinfo {author} {\bibfnamefont {D.-N.}\ \bibnamefont {Le}},\ and\ \bibinfo {author} {\bibfnamefont {L.~M.}\ \bibnamefont {Woods}},\ }\bibfield  {title} {\bibinfo {title} {Twisting in $h$-{BN} bilayers and their angle-dependent properties},\ }\href {https://doi.org/10.1103/wzz2-pszx} {\bibfield  {journal} {\bibinfo  {journal} {Phys. Rev. B}\ }\textbf {\bibinfo {volume} {112}},\ \bibinfo {pages} {085102} (\bibinfo {year} {2025})}\BibitemShut {NoStop}%
\bibitem [{\citenamefont {He}\ \emph {et~al.}(2024)\citenamefont {He}, \citenamefont {Zhang}, \citenamefont {Wang}, \citenamefont {Li}, \citenamefont {Tang}, \citenamefont {Bauer},\ and\ \citenamefont {Zhong}}]{He24p119416}%
  \BibitemOpen
  \bibfield  {author} {\bibinfo {author} {\bibfnamefont {R.}~\bibnamefont {He}}, \bibinfo {author} {\bibfnamefont {B.}~\bibnamefont {Zhang}}, \bibinfo {author} {\bibfnamefont {H.}~\bibnamefont {Wang}}, \bibinfo {author} {\bibfnamefont {L.}~\bibnamefont {Li}}, \bibinfo {author} {\bibfnamefont {P.}~\bibnamefont {Tang}}, \bibinfo {author} {\bibfnamefont {G.}~\bibnamefont {Bauer}},\ and\ \bibinfo {author} {\bibfnamefont {Z.}~\bibnamefont {Zhong}},\ }\bibfield  {title} {\bibinfo {title} {Ultrafast switching dynamics of the ferroelectric order in stacking-engineered ferroelectrics},\ }\href {https://doi.org/10.1016/j.actamat.2023.119416} {\bibfield  {journal} {\bibinfo  {journal} {Acta Mater.}\ }\textbf {\bibinfo {volume} {262}},\ \bibinfo {pages} {119416} (\bibinfo {year} {2024})}\BibitemShut {NoStop}%
\bibitem [{\citenamefont {Ko}\ \emph {et~al.}(2023)\citenamefont {Ko}, \citenamefont {Yuk}, \citenamefont {Engelke}, \citenamefont {Carr}, \citenamefont {Kim}, \citenamefont {Park}, \citenamefont {Heo}, \citenamefont {Kim}, \citenamefont {Kim}, \citenamefont {Kim}, \citenamefont {Taniguchi}, \citenamefont {Watanabe}, \citenamefont {Park}, \citenamefont {Kaxiras}, \citenamefont {Yang}, \citenamefont {Kim},\ and\ \citenamefont {Yoo}}]{Ko23p992}%
  \BibitemOpen
  \bibfield  {author} {\bibinfo {author} {\bibfnamefont {K.}~\bibnamefont {Ko}}, \bibinfo {author} {\bibfnamefont {A.}~\bibnamefont {Yuk}}, \bibinfo {author} {\bibfnamefont {R.}~\bibnamefont {Engelke}}, \bibinfo {author} {\bibfnamefont {S.}~\bibnamefont {Carr}}, \bibinfo {author} {\bibfnamefont {J.}~\bibnamefont {Kim}}, \bibinfo {author} {\bibfnamefont {D.}~\bibnamefont {Park}}, \bibinfo {author} {\bibfnamefont {H.}~\bibnamefont {Heo}}, \bibinfo {author} {\bibfnamefont {H.-M.}\ \bibnamefont {Kim}}, \bibinfo {author} {\bibfnamefont {S.-G.}\ \bibnamefont {Kim}}, \bibinfo {author} {\bibfnamefont {H.}~\bibnamefont {Kim}}, \bibinfo {author} {\bibfnamefont {T.}~\bibnamefont {Taniguchi}}, \bibinfo {author} {\bibfnamefont {K.}~\bibnamefont {Watanabe}}, \bibinfo {author} {\bibfnamefont {H.}~\bibnamefont {Park}}, \bibinfo {author} {\bibfnamefont {E.}~\bibnamefont {Kaxiras}}, \bibinfo {author} {\bibfnamefont {S.~M.}\ \bibnamefont {Yang}}, \bibinfo {author} {\bibfnamefont {P.}~\bibnamefont {Kim}},\ and\ \bibinfo {author}
  {\bibfnamefont {H.}~\bibnamefont {Yoo}},\ }\bibfield  {title} {\bibinfo {title} {Operando electron microscopy investigation of polar domain dynamics in twisted van der {Waals} homobilayers},\ }\href {https://doi.org/10.1038/s41563-023-01595-0} {\bibfield  {journal} {\bibinfo  {journal} {Nat. Mater.}\ }\textbf {\bibinfo {volume} {22}},\ \bibinfo {pages} {992} (\bibinfo {year} {2023})}\BibitemShut {NoStop}%
\bibitem [{\citenamefont {Wong}\ \emph {et~al.}(2025)\citenamefont {Wong}, \citenamefont {Lin}, \citenamefont {Ho}, \citenamefont {Hsu}, \citenamefont {Li}, \citenamefont {Chen}, \citenamefont {Huang}, \citenamefont {Chang}, \citenamefont {Hsieh}, \citenamefont {Chen} \emph {et~al.}}]{wong25p2414442}%
  \BibitemOpen
  \bibfield  {author} {\bibinfo {author} {\bibfnamefont {S.-S.}\ \bibnamefont {Wong}}, \bibinfo {author} {\bibfnamefont {Z.-Y.}\ \bibnamefont {Lin}}, \bibinfo {author} {\bibfnamefont {S.-Z.}\ \bibnamefont {Ho}}, \bibinfo {author} {\bibfnamefont {C.-E.}\ \bibnamefont {Hsu}}, \bibinfo {author} {\bibfnamefont {P.-H.}\ \bibnamefont {Li}}, \bibinfo {author} {\bibfnamefont {C.-Y.}\ \bibnamefont {Chen}}, \bibinfo {author} {\bibfnamefont {Y.-F.}\ \bibnamefont {Huang}}, \bibinfo {author} {\bibfnamefont {K.-E.}\ \bibnamefont {Chang}}, \bibinfo {author} {\bibfnamefont {Y.-C.}\ \bibnamefont {Hsieh}}, \bibinfo {author} {\bibfnamefont {C.-H.}\ \bibnamefont {Chen}}, \emph {et~al.},\ }\bibfield  {title} {\bibinfo {title} {Epitaxial ferroelectric hexagonal boron nitride grown on graphene},\ }\href {https://doi.org/10.1002/adma.202414442} {\bibfield  {journal} {\bibinfo  {journal} {Adv. Mater.}\ }\textbf {\bibinfo {volume} {37}},\ \bibinfo {pages} {2414442} (\bibinfo {year} {2025})}\BibitemShut {NoStop}%
\bibitem [{\citenamefont {Fan}\ \emph {et~al.}(2025)\citenamefont {Fan}, \citenamefont {Guan}, \citenamefont {Wei}, \citenamefont {Xu}, \citenamefont {Tong}, \citenamefont {Tian}, \citenamefont {Wan}, \citenamefont {Yao}, \citenamefont {Zheng}, \citenamefont {Chen} \emph {et~al.}}]{fan25p3557}%
  \BibitemOpen
  \bibfield  {author} {\bibinfo {author} {\bibfnamefont {W.-C.}\ \bibnamefont {Fan}}, \bibinfo {author} {\bibfnamefont {Z.}~\bibnamefont {Guan}}, \bibinfo {author} {\bibfnamefont {L.-Q.}\ \bibnamefont {Wei}}, \bibinfo {author} {\bibfnamefont {H.-W.}\ \bibnamefont {Xu}}, \bibinfo {author} {\bibfnamefont {W.-Y.}\ \bibnamefont {Tong}}, \bibinfo {author} {\bibfnamefont {M.}~\bibnamefont {Tian}}, \bibinfo {author} {\bibfnamefont {N.}~\bibnamefont {Wan}}, \bibinfo {author} {\bibfnamefont {C.-S.}\ \bibnamefont {Yao}}, \bibinfo {author} {\bibfnamefont {J.-D.}\ \bibnamefont {Zheng}}, \bibinfo {author} {\bibfnamefont {B.-B.}\ \bibnamefont {Chen}}, \emph {et~al.},\ }\bibfield  {title} {\bibinfo {title} {Edge polarization topology integrated with sliding ferroelectricity in moir{\'e} system},\ }\href {https://doi.org/10.1038/s41467-025-58877-1} {\bibfield  {journal} {\bibinfo  {journal} {Nat. Commun.}\ }\textbf {\bibinfo {volume} {16}},\ \bibinfo {pages} {3557} (\bibinfo {year} {2025})}\BibitemShut {NoStop}%
\bibitem [{\citenamefont {Lv}\ \emph {et~al.}(2022)\citenamefont {Lv}, \citenamefont {Sun}, \citenamefont {Chen}, \citenamefont {Taniguchi}, \citenamefont {Watanabe}, \citenamefont {Wu}, \citenamefont {Wang},\ and\ \citenamefont {Xue}}]{ming22p21}%
  \BibitemOpen
  \bibfield  {author} {\bibinfo {author} {\bibfnamefont {M.}~\bibnamefont {Lv}}, \bibinfo {author} {\bibfnamefont {X.}~\bibnamefont {Sun}}, \bibinfo {author} {\bibfnamefont {Y.}~\bibnamefont {Chen}}, \bibinfo {author} {\bibfnamefont {T.}~\bibnamefont {Taniguchi}}, \bibinfo {author} {\bibfnamefont {K.}~\bibnamefont {Watanabe}}, \bibinfo {author} {\bibfnamefont {M.}~\bibnamefont {Wu}}, \bibinfo {author} {\bibfnamefont {J.}~\bibnamefont {Wang}},\ and\ \bibinfo {author} {\bibfnamefont {J.}~\bibnamefont {Xue}},\ }\bibfield  {title} {\bibinfo {title} {Spatially resolved polarization manipulation of ferroelectricity in twisted {h-BN}},\ }\href {https://doi.org/10.1002/adma.202203990} {\bibfield  {journal} {\bibinfo  {journal} {Adv. Mater.}\ }\textbf {\bibinfo {volume} {34}},\ \bibinfo {pages} {2203990} (\bibinfo {year} {2022})}\BibitemShut {NoStop}%
\bibitem [{\citenamefont {Tybell}\ \emph {et~al.}(2002)\citenamefont {Tybell}, \citenamefont {Paruch}, \citenamefont {Giamarchi},\ and\ \citenamefont {Triscone}}]{Tybell02p097601}%
  \BibitemOpen
  \bibfield  {author} {\bibinfo {author} {\bibfnamefont {T.}~\bibnamefont {Tybell}}, \bibinfo {author} {\bibfnamefont {P.}~\bibnamefont {Paruch}}, \bibinfo {author} {\bibfnamefont {T.}~\bibnamefont {Giamarchi}},\ and\ \bibinfo {author} {\bibfnamefont {J.}~\bibnamefont {Triscone}},\ }\bibfield  {title} {\bibinfo {title} {Domain wall creep in epitaxial ferroelectric {Pb(Zr$_{0.2}$Ti$_{0.8}$)O$_3$} thin films},\ }\href {https://doi.org/10.1103/physrevlett.89.097601} {\bibfield  {journal} {\bibinfo  {journal} {Phys. Rev. Lett.}\ }\textbf {\bibinfo {volume} {89}},\ \bibinfo {pages} {097601} (\bibinfo {year} {2002})}\BibitemShut {NoStop}%
\bibitem [{\citenamefont {Li}\ \emph {et~al.}(2004)\citenamefont {Li}, \citenamefont {Wang}, \citenamefont {Wuttig}, \citenamefont {Ramesh}, \citenamefont {Wang}, \citenamefont {Ruette}, \citenamefont {Pyatakov}, \citenamefont {Zvezdin},\ and\ \citenamefont {Viehland}}]{Li04p5261}%
  \BibitemOpen
  \bibfield  {author} {\bibinfo {author} {\bibfnamefont {J.}~\bibnamefont {Li}}, \bibinfo {author} {\bibfnamefont {J.}~\bibnamefont {Wang}}, \bibinfo {author} {\bibfnamefont {M.}~\bibnamefont {Wuttig}}, \bibinfo {author} {\bibfnamefont {R.}~\bibnamefont {Ramesh}}, \bibinfo {author} {\bibfnamefont {N.}~\bibnamefont {Wang}}, \bibinfo {author} {\bibfnamefont {B.}~\bibnamefont {Ruette}}, \bibinfo {author} {\bibfnamefont {A.~P.}\ \bibnamefont {Pyatakov}}, \bibinfo {author} {\bibfnamefont {A.}~\bibnamefont {Zvezdin}},\ and\ \bibinfo {author} {\bibfnamefont {D.}~\bibnamefont {Viehland}},\ }\bibfield  {title} {\bibinfo {title} {Dramatically enhanced polarization in (001), (101), and (111) {BiFeO}$_3$ thin films due to epitiaxial-induced transitions},\ }\href {https://doi.org/10.1063/1.1764944} {\bibfield  {journal} {\bibinfo  {journal} {Appl. Phys. Lett.}\ }\textbf {\bibinfo {volume} {84}},\ \bibinfo {pages} {5261} (\bibinfo {year} {2004})}\BibitemShut {NoStop}%
\bibitem [{\citenamefont {T\"uckmantel}\ \emph {et~al.}(2021)\citenamefont {T\"uckmantel}, \citenamefont {Gaponenko}, \citenamefont {Caballero}, \citenamefont {Agar}, \citenamefont {Martin}, \citenamefont {Giamarchi},\ and\ \citenamefont {Paruch}}]{philippe21p117601}%
  \BibitemOpen
  \bibfield  {author} {\bibinfo {author} {\bibfnamefont {P.}~\bibnamefont {T\"uckmantel}}, \bibinfo {author} {\bibfnamefont {I.}~\bibnamefont {Gaponenko}}, \bibinfo {author} {\bibfnamefont {N.}~\bibnamefont {Caballero}}, \bibinfo {author} {\bibfnamefont {J.~C.}\ \bibnamefont {Agar}}, \bibinfo {author} {\bibfnamefont {L.~W.}\ \bibnamefont {Martin}}, \bibinfo {author} {\bibfnamefont {T.}~\bibnamefont {Giamarchi}},\ and\ \bibinfo {author} {\bibfnamefont {P.}~\bibnamefont {Paruch}},\ }\bibfield  {title} {\bibinfo {title} {Local probe comparison of ferroelectric switching event statistics in the creep and depinning regimes in {Pb(Zr$_{0.2}$Ti$_{0.8}$)O$_3$} thin films},\ }\href {https://doi.org/10.1103/PhysRevLett.126.117601} {\bibfield  {journal} {\bibinfo  {journal} {Phys. Rev. Lett.}\ }\textbf {\bibinfo {volume} {126}},\ \bibinfo {pages} {117601} (\bibinfo {year} {2021})}\BibitemShut {NoStop}%
\bibitem [{\citenamefont {Li}\ \emph {et~al.}(2025)\citenamefont {Li}, \citenamefont {Wei}, \citenamefont {Guo}, \citenamefont {Wang}, \citenamefont {Zhang}, \citenamefont {Taniguchi}, \citenamefont {Watanabe}, \citenamefont {Shi}, \citenamefont {Shi}, \citenamefont {Wang} \emph {et~al.}}]{li25p5451}%
  \BibitemOpen
  \bibfield  {author} {\bibinfo {author} {\bibfnamefont {Y.}~\bibnamefont {Li}}, \bibinfo {author} {\bibfnamefont {Y.}~\bibnamefont {Wei}}, \bibinfo {author} {\bibfnamefont {R.}~\bibnamefont {Guo}}, \bibinfo {author} {\bibfnamefont {Y.}~\bibnamefont {Wang}}, \bibinfo {author} {\bibfnamefont {H.}~\bibnamefont {Zhang}}, \bibinfo {author} {\bibfnamefont {T.}~\bibnamefont {Taniguchi}}, \bibinfo {author} {\bibfnamefont {K.}~\bibnamefont {Watanabe}}, \bibinfo {author} {\bibfnamefont {Y.}~\bibnamefont {Shi}}, \bibinfo {author} {\bibfnamefont {Y.}~\bibnamefont {Shi}}, \bibinfo {author} {\bibfnamefont {C.}~\bibnamefont {Wang}}, \emph {et~al.},\ }\bibfield  {title} {\bibinfo {title} {Unusual topological polar texture in moir{\'e} ferroelectrics},\ }\href {https://doi.org/10.1038/s41467-025-60647-y} {\bibfield  {journal} {\bibinfo  {journal} {Nat. Commun.}\ }\textbf {\bibinfo {volume} {16}},\ \bibinfo {pages} {5451} (\bibinfo {year} {2025})}\BibitemShut {NoStop}%
\bibitem [{\citenamefont {Weston}\ \emph {et~al.}(2018)\citenamefont {Weston}, \citenamefont {Wickramaratne}, \citenamefont {Mackoit}, \citenamefont {Alkauskas},\ and\ \citenamefont {Van~de Walle}}]{weston20p214104}%
  \BibitemOpen
  \bibfield  {author} {\bibinfo {author} {\bibfnamefont {L.}~\bibnamefont {Weston}}, \bibinfo {author} {\bibfnamefont {D.}~\bibnamefont {Wickramaratne}}, \bibinfo {author} {\bibfnamefont {M.}~\bibnamefont {Mackoit}}, \bibinfo {author} {\bibfnamefont {A.}~\bibnamefont {Alkauskas}},\ and\ \bibinfo {author} {\bibfnamefont {C.~G.}\ \bibnamefont {Van~de Walle}},\ }\bibfield  {title} {\bibinfo {title} {Native point defects and impurities in hexagonal boron nitride},\ }\href {https://doi.org/10.1103/PhysRevB.97.214104} {\bibfield  {journal} {\bibinfo  {journal} {Phys. Rev. B}\ }\textbf {\bibinfo {volume} {97}},\ \bibinfo {pages} {214104} (\bibinfo {year} {2018})}\BibitemShut {NoStop}%
\bibitem [{\citenamefont {Uddin}\ \emph {et~al.}(2017)\citenamefont {Uddin}, \citenamefont {Li}, \citenamefont {Lin},\ and\ \citenamefont {Jiang}}]{uddin17p18}%
  \BibitemOpen
  \bibfield  {author} {\bibinfo {author} {\bibfnamefont {M.}~\bibnamefont {Uddin}}, \bibinfo {author} {\bibfnamefont {J.}~\bibnamefont {Li}}, \bibinfo {author} {\bibfnamefont {J.}~\bibnamefont {Lin}},\ and\ \bibinfo {author} {\bibfnamefont {H.}~\bibnamefont {Jiang}},\ }\bibfield  {title} {\bibinfo {title} {Probing carbon impurities in hexagonal boron nitride epilayers},\ }\href {https://doi.org/10.1063/1.4982647} {\bibfield  {journal} {\bibinfo  {journal} {Appl. Phys. Lett.}\ }\textbf {\bibinfo {volume} {110}},\ \bibinfo {pages} {182107} (\bibinfo {year} {2017})}\BibitemShut {NoStop}%
\bibitem [{\citenamefont {Kim}\ \emph {et~al.}(2023)\citenamefont {Kim}, \citenamefont {Haque}, \citenamefont {Hsieh}, \citenamefont {Nahid}, \citenamefont {Zarin}, \citenamefont {Jeong}, \citenamefont {So}, \citenamefont {Park},\ and\ \citenamefont {Nam}}]{kim23p2107362}%
  \BibitemOpen
  \bibfield  {author} {\bibinfo {author} {\bibfnamefont {J.~M.}\ \bibnamefont {Kim}}, \bibinfo {author} {\bibfnamefont {M.~F.}\ \bibnamefont {Haque}}, \bibinfo {author} {\bibfnamefont {E.~Y.}\ \bibnamefont {Hsieh}}, \bibinfo {author} {\bibfnamefont {S.~M.}\ \bibnamefont {Nahid}}, \bibinfo {author} {\bibfnamefont {I.}~\bibnamefont {Zarin}}, \bibinfo {author} {\bibfnamefont {K.-Y.}\ \bibnamefont {Jeong}}, \bibinfo {author} {\bibfnamefont {J.-P.}\ \bibnamefont {So}}, \bibinfo {author} {\bibfnamefont {H.-G.}\ \bibnamefont {Park}},\ and\ \bibinfo {author} {\bibfnamefont {S.}~\bibnamefont {Nam}},\ }\bibfield  {title} {\bibinfo {title} {Strain engineering of low-dimensional materials for emerging quantum phenomena and functionalities},\ }\href {https://doi.org/10.1002/adma.202107362} {\bibfield  {journal} {\bibinfo  {journal} {Adv. Mater.}\ }\textbf {\bibinfo {volume} {35}},\ \bibinfo {pages} {2107362} (\bibinfo {year} {2023})}\BibitemShut {NoStop}%
\bibitem [{\citenamefont {Blundo}\ \emph {et~al.}(2022)\citenamefont {Blundo}, \citenamefont {Surrente}, \citenamefont {Spirito}, \citenamefont {Pettinari}, \citenamefont {Yildirim}, \citenamefont {Chavarin}, \citenamefont {Baldassarre}, \citenamefont {Felici},\ and\ \citenamefont {Polimeni}}]{Blundo22p1525}%
  \BibitemOpen
  \bibfield  {author} {\bibinfo {author} {\bibfnamefont {E.}~\bibnamefont {Blundo}}, \bibinfo {author} {\bibfnamefont {A.}~\bibnamefont {Surrente}}, \bibinfo {author} {\bibfnamefont {D.}~\bibnamefont {Spirito}}, \bibinfo {author} {\bibfnamefont {G.}~\bibnamefont {Pettinari}}, \bibinfo {author} {\bibfnamefont {T.}~\bibnamefont {Yildirim}}, \bibinfo {author} {\bibfnamefont {C.~A.}\ \bibnamefont {Chavarin}}, \bibinfo {author} {\bibfnamefont {L.}~\bibnamefont {Baldassarre}}, \bibinfo {author} {\bibfnamefont {M.}~\bibnamefont {Felici}},\ and\ \bibinfo {author} {\bibfnamefont {A.}~\bibnamefont {Polimeni}},\ }\bibfield  {title} {\bibinfo {title} {Vibrational properties in highly strained hexagonal boron nitride bubbles},\ }\href {https://doi.org/10.1021/acs.nanolett.1c04197} {\bibfield  {journal} {\bibinfo  {journal} {Nano Lett.}\ }\textbf {\bibinfo {volume} {22}},\ \bibinfo {pages} {1525} (\bibinfo {year} {2022})}\BibitemShut {NoStop}%
\bibitem [{\citenamefont {Wang}\ \emph {et~al.}(2019)\citenamefont {Wang}, \citenamefont {Dai}, \citenamefont {Xiao}, \citenamefont {Feng}, \citenamefont {Weng}, \citenamefont {Liu}, \citenamefont {Xu}, \citenamefont {Huang},\ and\ \citenamefont {Zhang}}]{wang19p116101}%
  \BibitemOpen
  \bibfield  {author} {\bibinfo {author} {\bibfnamefont {G.}~\bibnamefont {Wang}}, \bibinfo {author} {\bibfnamefont {Z.}~\bibnamefont {Dai}}, \bibinfo {author} {\bibfnamefont {J.}~\bibnamefont {Xiao}}, \bibinfo {author} {\bibfnamefont {S.}~\bibnamefont {Feng}}, \bibinfo {author} {\bibfnamefont {C.}~\bibnamefont {Weng}}, \bibinfo {author} {\bibfnamefont {L.}~\bibnamefont {Liu}}, \bibinfo {author} {\bibfnamefont {Z.}~\bibnamefont {Xu}}, \bibinfo {author} {\bibfnamefont {R.}~\bibnamefont {Huang}},\ and\ \bibinfo {author} {\bibfnamefont {Z.}~\bibnamefont {Zhang}},\ }\bibfield  {title} {\bibinfo {title} {Bending of multilayer van der waals materials},\ }\href {https://doi.org/10.1103/PhysRevLett.123.116101} {\bibfield  {journal} {\bibinfo  {journal} {Phys. Rev. Lett.}\ }\textbf {\bibinfo {volume} {123}},\ \bibinfo {pages} {116101} (\bibinfo {year} {2019})}\BibitemShut {NoStop}%
\bibitem [{\citenamefont {Zhang}\ \emph {et~al.}(2018{\natexlab{a}})\citenamefont {Zhang}, \citenamefont {Han}, \citenamefont {Wang}, \citenamefont {Car},\ and\ \citenamefont {E}}]{Zhang18p143001}%
  \BibitemOpen
  \bibfield  {author} {\bibinfo {author} {\bibfnamefont {L.}~\bibnamefont {Zhang}}, \bibinfo {author} {\bibfnamefont {J.}~\bibnamefont {Han}}, \bibinfo {author} {\bibfnamefont {H.}~\bibnamefont {Wang}}, \bibinfo {author} {\bibfnamefont {R.}~\bibnamefont {Car}},\ and\ \bibinfo {author} {\bibfnamefont {W.}~\bibnamefont {E}},\ }\bibfield  {title} {\bibinfo {title} {Deep potential molecular dynamics: A scalable model with the accuracy of quantum mechanics},\ }\href {https://doi.org/10.1103/physrevlett.120.143001} {\bibfield  {journal} {\bibinfo  {journal} {Phys. Rev. Lett.}\ }\textbf {\bibinfo {volume} {120}},\ \bibinfo {pages} {143001} (\bibinfo {year} {2018}{\natexlab{a}})}\BibitemShut {NoStop}%
\bibitem [{\citenamefont {Zhang}\ \emph {et~al.}(2018{\natexlab{b}})\citenamefont {Zhang}, \citenamefont {Han}, \citenamefont {Wang}, \citenamefont {Saidi}, \citenamefont {Car},\ and\ \citenamefont {Weinan}}]{Zhang18p4441}%
  \BibitemOpen
  \bibfield  {author} {\bibinfo {author} {\bibfnamefont {L.}~\bibnamefont {Zhang}}, \bibinfo {author} {\bibfnamefont {J.}~\bibnamefont {Han}}, \bibinfo {author} {\bibfnamefont {H.}~\bibnamefont {Wang}}, \bibinfo {author} {\bibfnamefont {W.~A.}\ \bibnamefont {Saidi}}, \bibinfo {author} {\bibfnamefont {R.}~\bibnamefont {Car}},\ and\ \bibinfo {author} {\bibfnamefont {E.}~\bibnamefont {Weinan}},\ }\bibfield  {title} {\bibinfo {title} {End-to-end symmetry preserving inter-atomic potential energy model for finite and extended systems},\ }in\ \href {https://doi.org/10.1016/j.cpc.2020.107206} {\emph {\bibinfo {booktitle} {Proceedings of the 32nd International Conference on Neural Information Processing Systems}}},\ \bibinfo {series and number} {NIPS'18}\ (\bibinfo  {publisher} {Curran Associates Inc.},\ \bibinfo {address} {Red Hook, NY, USA},\ \bibinfo {year} {2018})\ pp.\ \bibinfo {pages} {4441--4451}\BibitemShut {NoStop}%
\bibitem [{\citenamefont {Ke}\ \emph {et~al.}(2025)\citenamefont {Ke}, \citenamefont {Liu},\ and\ \citenamefont {Liu}}]{ke25p046201}%
  \BibitemOpen
  \bibfield  {author} {\bibinfo {author} {\bibfnamefont {C.}~\bibnamefont {Ke}}, \bibinfo {author} {\bibfnamefont {F.}~\bibnamefont {Liu}},\ and\ \bibinfo {author} {\bibfnamefont {S.}~\bibnamefont {Liu}},\ }\bibfield  {title} {\bibinfo {title} {Superlubric motion of wavelike domain walls in sliding ferroelectrics},\ }\href {https://doi.org/10.1103/jhlq-2dd7} {\bibfield  {journal} {\bibinfo  {journal} {Phys. Rev. Lett.}\ }\textbf {\bibinfo {volume} {135}},\ \bibinfo {pages} {046201} (\bibinfo {year} {2025})}\BibitemShut {NoStop}%
\bibitem [{\citenamefont {Durdiev}\ \emph {et~al.}(2024)\citenamefont {Durdiev}, \citenamefont {Zaiser}, \citenamefont {Wendler}, \citenamefont {Tsuzuki}, \citenamefont {Azuma}, \citenamefont {Ogata}, \citenamefont {Kobayashi},\ and\ \citenamefont {Uranagase}}]{dilshood24p132901}%
  \BibitemOpen
  \bibfield  {author} {\bibinfo {author} {\bibfnamefont {D.}~\bibnamefont {Durdiev}}, \bibinfo {author} {\bibfnamefont {M.}~\bibnamefont {Zaiser}}, \bibinfo {author} {\bibfnamefont {F.}~\bibnamefont {Wendler}}, \bibinfo {author} {\bibfnamefont {T.}~\bibnamefont {Tsuzuki}}, \bibinfo {author} {\bibfnamefont {H.}~\bibnamefont {Azuma}}, \bibinfo {author} {\bibfnamefont {S.}~\bibnamefont {Ogata}}, \bibinfo {author} {\bibfnamefont {R.}~\bibnamefont {Kobayashi}},\ and\ \bibinfo {author} {\bibfnamefont {M.}~\bibnamefont {Uranagase}},\ }\bibfield  {title} {\bibinfo {title} {Determining thermal activation parameters for ferroelectric domain nucleation in {BaTiO$_3$} from molecular dynamics simulations},\ }\href {https://doi.org/10.1063/5.0187476} {\bibfield  {journal} {\bibinfo  {journal} {Appl. Phys. Lett.}\ }\textbf {\bibinfo {volume} {124}},\ \bibinfo {pages} {132901} (\bibinfo {year} {2024})}\BibitemShut {NoStop}%
\bibitem [{\citenamefont {Sm\aa{}br\aa{}ten}\ \emph {et~al.}(2020)\citenamefont {Sm\aa{}br\aa{}ten}, \citenamefont {Holstad}, \citenamefont {Evans}, \citenamefont {Yan}, \citenamefont {Bourret}, \citenamefont {Meier},\ and\ \citenamefont {Selbach}}]{dr20p033159}%
  \BibitemOpen
  \bibfield  {author} {\bibinfo {author} {\bibfnamefont {D.~R.}\ \bibnamefont {Sm\aa{}br\aa{}ten}}, \bibinfo {author} {\bibfnamefont {T.~S.}\ \bibnamefont {Holstad}}, \bibinfo {author} {\bibfnamefont {D.~M.}\ \bibnamefont {Evans}}, \bibinfo {author} {\bibfnamefont {Z.}~\bibnamefont {Yan}}, \bibinfo {author} {\bibfnamefont {E.}~\bibnamefont {Bourret}}, \bibinfo {author} {\bibfnamefont {D.}~\bibnamefont {Meier}},\ and\ \bibinfo {author} {\bibfnamefont {S.~M.}\ \bibnamefont {Selbach}},\ }\bibfield  {title} {\bibinfo {title} {Domain wall mobility and roughening in doped ferroelectric hexagonal manganites},\ }\href {https://doi.org/10.1103/PhysRevResearch.2.033159} {\bibfield  {journal} {\bibinfo  {journal} {Phys. Rev. Res.}\ }\textbf {\bibinfo {volume} {2}},\ \bibinfo {pages} {033159} (\bibinfo {year} {2020})}\BibitemShut {NoStop}%
\bibitem [{\citenamefont {Bencan}\ \emph {et~al.}(2020)\citenamefont {Bencan}, \citenamefont {Drazic}, \citenamefont {Ursic}, \citenamefont {Makarovic}, \citenamefont {Komelj},\ and\ \citenamefont {Rojac}}]{bencan20p1762}%
  \BibitemOpen
  \bibfield  {author} {\bibinfo {author} {\bibfnamefont {A.}~\bibnamefont {Bencan}}, \bibinfo {author} {\bibfnamefont {G.}~\bibnamefont {Drazic}}, \bibinfo {author} {\bibfnamefont {H.}~\bibnamefont {Ursic}}, \bibinfo {author} {\bibfnamefont {M.}~\bibnamefont {Makarovic}}, \bibinfo {author} {\bibfnamefont {M.}~\bibnamefont {Komelj}},\ and\ \bibinfo {author} {\bibfnamefont {T.}~\bibnamefont {Rojac}},\ }\bibfield  {title} {\bibinfo {title} {Domain-wall pinning and defect ordering in {BiFeO$_3$} probed on the atomic and nanoscale},\ }\href {https://doi.org/10.1038/s41467-020-15595-0} {\bibfield  {journal} {\bibinfo  {journal} {Nat. Commun.}\ }\textbf {\bibinfo {volume} {11}},\ \bibinfo {pages} {1762} (\bibinfo {year} {2020})}\BibitemShut {NoStop}%
\bibitem [{\citenamefont {Wang}\ and\ \citenamefont {Dong}(2025)}]{dong25pL201406}%
  \BibitemOpen
  \bibfield  {author} {\bibinfo {author} {\bibfnamefont {Z.}~\bibnamefont {Wang}}\ and\ \bibinfo {author} {\bibfnamefont {S.}~\bibnamefont {Dong}},\ }\bibfield  {title} {\bibinfo {title} {Polarization switching in sliding ferroelectrics: Roles of fluctuation and domain wall},\ }\href {https://doi.org/10.1103/PhysRevB.111.L201406} {\bibfield  {journal} {\bibinfo  {journal} {Phys. Rev. B}\ }\textbf {\bibinfo {volume} {111}},\ \bibinfo {pages} {L201406} (\bibinfo {year} {2025})}\BibitemShut {NoStop}%
\bibitem [{\citenamefont {Xu}\ \emph {et~al.}(2025)\citenamefont {Xu}, \citenamefont {Fan}, \citenamefont {Zheng}, \citenamefont {Yao}, \citenamefont {Zhong}, \citenamefont {Tong},\ and\ \citenamefont {Duan}}]{xu25parXiv}%
  \BibitemOpen
  \bibfield  {author} {\bibinfo {author} {\bibfnamefont {H.-W.}\ \bibnamefont {Xu}}, \bibinfo {author} {\bibfnamefont {W.-C.}\ \bibnamefont {Fan}}, \bibinfo {author} {\bibfnamefont {J.-D.}\ \bibnamefont {Zheng}}, \bibinfo {author} {\bibfnamefont {C.-S.}\ \bibnamefont {Yao}}, \bibinfo {author} {\bibfnamefont {N.}~\bibnamefont {Zhong}}, \bibinfo {author} {\bibfnamefont {W.-Y.}\ \bibnamefont {Tong}},\ and\ \bibinfo {author} {\bibfnamefont {C.-G.}\ \bibnamefont {Duan}},\ }\href {https://doi.org/10.48550/arXiv.2508.11997} {\bibinfo {title} {Domain wall-mediated interfacial ferroelectric switching}} (\bibinfo {year} {2025}),\ \Eprint {https://arxiv.org/abs/2508.11997} {arXiv:2508.11997} \BibitemShut {NoStop}%
\bibitem [{\citenamefont {Shi}\ \emph {et~al.}(2025)\citenamefont {Shi}, \citenamefont {Gao}, \citenamefont {Wang}, \citenamefont {Zhang}, \citenamefont {Zhong},\ and\ \citenamefont {He}}]{shi25p035421}%
  \BibitemOpen
  \bibfield  {author} {\bibinfo {author} {\bibfnamefont {Y.}~\bibnamefont {Shi}}, \bibinfo {author} {\bibfnamefont {Y.}~\bibnamefont {Gao}}, \bibinfo {author} {\bibfnamefont {H.}~\bibnamefont {Wang}}, \bibinfo {author} {\bibfnamefont {B.}~\bibnamefont {Zhang}}, \bibinfo {author} {\bibfnamefont {Z.}~\bibnamefont {Zhong}},\ and\ \bibinfo {author} {\bibfnamefont {R.}~\bibnamefont {He}},\ }\bibfield  {title} {\bibinfo {title} {Soliton-like domain wall motion in sliding ferroelectrics with ultralow damping},\ }\href {https://doi.org/10.1103/b91v-r2rc} {\bibfield  {journal} {\bibinfo  {journal} {Phys. Rev. B}\ }\textbf {\bibinfo {volume} {112}},\ \bibinfo {pages} {035421} (\bibinfo {year} {2025})}\BibitemShut {NoStop}%
\bibitem [{\citenamefont {Bian}\ \emph {et~al.}(2024)\citenamefont {Bian}, \citenamefont {He}, \citenamefont {Pan}, \citenamefont {Li}, \citenamefont {Cao}, \citenamefont {Meng}, \citenamefont {Chen}, \citenamefont {Liu}, \citenamefont {Zhong}, \citenamefont {Li},\ and\ \citenamefont {Liu}}]{Bian24p57}%
  \BibitemOpen
  \bibfield  {author} {\bibinfo {author} {\bibfnamefont {R.}~\bibnamefont {Bian}}, \bibinfo {author} {\bibfnamefont {R.}~\bibnamefont {He}}, \bibinfo {author} {\bibfnamefont {E.}~\bibnamefont {Pan}}, \bibinfo {author} {\bibfnamefont {Z.}~\bibnamefont {Li}}, \bibinfo {author} {\bibfnamefont {G.}~\bibnamefont {Cao}}, \bibinfo {author} {\bibfnamefont {P.}~\bibnamefont {Meng}}, \bibinfo {author} {\bibfnamefont {J.}~\bibnamefont {Chen}}, \bibinfo {author} {\bibfnamefont {Q.}~\bibnamefont {Liu}}, \bibinfo {author} {\bibfnamefont {Z.}~\bibnamefont {Zhong}}, \bibinfo {author} {\bibfnamefont {W.}~\bibnamefont {Li}},\ and\ \bibinfo {author} {\bibfnamefont {F.}~\bibnamefont {Liu}},\ }\bibfield  {title} {\bibinfo {title} {Developing fatigue-resistant ferroelectrics using interlayer sliding switching},\ }\href {https://doi.org/10.1126/science.ado1744} {\bibfield  {journal} {\bibinfo  {journal} {Science}\ }\textbf {\bibinfo {volume} {385}},\ \bibinfo {pages} {57} (\bibinfo {year} {2024})}\BibitemShut {NoStop}%
\bibitem [{\citenamefont {Yasuda}\ \emph {et~al.}(2024)\citenamefont {Yasuda}, \citenamefont {Zalys-Geller}, \citenamefont {Wang}, \citenamefont {Bennett}, \citenamefont {Cheema}, \citenamefont {Watanabe}, \citenamefont {Taniguchi}, \citenamefont {Kaxiras}, \citenamefont {Jarillo-Herrero},\ and\ \citenamefont {Ashoori}}]{Yasuda24peadp3575}%
  \BibitemOpen
  \bibfield  {author} {\bibinfo {author} {\bibfnamefont {K.}~\bibnamefont {Yasuda}}, \bibinfo {author} {\bibfnamefont {E.}~\bibnamefont {Zalys-Geller}}, \bibinfo {author} {\bibfnamefont {X.}~\bibnamefont {Wang}}, \bibinfo {author} {\bibfnamefont {D.}~\bibnamefont {Bennett}}, \bibinfo {author} {\bibfnamefont {S.~S.}\ \bibnamefont {Cheema}}, \bibinfo {author} {\bibfnamefont {K.}~\bibnamefont {Watanabe}}, \bibinfo {author} {\bibfnamefont {T.}~\bibnamefont {Taniguchi}}, \bibinfo {author} {\bibfnamefont {E.}~\bibnamefont {Kaxiras}}, \bibinfo {author} {\bibfnamefont {P.}~\bibnamefont {Jarillo-Herrero}},\ and\ \bibinfo {author} {\bibfnamefont {R.}~\bibnamefont {Ashoori}},\ }\bibfield  {title} {\bibinfo {title} {Ultrafast high-endurance memory based on sliding ferroelectrics},\ }\href {https://doi.org/10.1126/science.adp3575} {\bibfield  {journal} {\bibinfo  {journal} {Science}\ }\textbf {\bibinfo {volume} {385}},\ \bibinfo {pages} {eadp3575} (\bibinfo {year} {2024})}\BibitemShut {NoStop}%
\end{thebibliography}%

\clearpage

\clearpage

\clearpage

\clearpage

\clearpage

\end{document}